\documentclass[12pt, letterpaper]{article}
\usepackage[utf8]{inputenc}
\usepackage{float,cite,graphicx,amsmath,amsthm,amssymb,subcaption,caption,bbold,amstext,array,authblk}
\usepackage{epstopdf}
\usepackage[nottoc]{tocbibind}
\newcolumntype{L}{>{$}l<{$}} % math-mode version of "l" column type
\usepackage[margin=0.8in]{geometry}
\usepackage{setspace}
\newcommand{\bom}{\boldsymbol{\omega}}
\newcommand{\bphi}{\boldsymbol{\Phi}}
\newcommand{\bv}{\mathbf{v}}
\newcommand{\bx}{\mathbf{x}}
\newcommand{\bu}{\mathbf{u}}
\newcommand{\bq}{\mathbf{q}}
\newcommand{\bp}{\mathbf{p}}
\newcommand{\by}{\mathbf{y}}
\newcommand{\bm}{\mathbf{m}}
\newcommand{\bL}{\mathbf{m}}

\newcommand{\ee}{\mathrm{e}}
\newcommand{\be}{\boldsymbol{e}}

\newcommand{\trans}{^{\mathrm{T}}}
\newcommand{\nh}{\nabla H}
\newcommand{\bd}{\boldsymbol{\delta}}

\begin{document}
\title{ \textbf{A novel approach to rigid spheroid models in viscous flows using operator splitting methods}}
\author[a]{Benjamin Tapley\thanks{Corresponding author. E-mail: \texttt{benjamin.tapley@ntnu.no}. Present address: Department of Mathematical Sciences, The Norwegian University of Science and Technology, 7491 Trondheim, Norway.}}
\author[a]{Elena Celledoni}
\author[a]{Brynjulf Owren}
\author[b]{Helge I. Andersson}
\affil[a]{Department of Mathematical Sciences, NTNU, Norway}
\affil[b]{Department of Energy and Process Engineering, NTNU, Norway}
\date{\today}
\maketitle

\begin{abstract}
Calculating cost-effective solutions to particle dynamics in viscous flows is an important problem in many areas of industry and nature.  We implement a second-order symmetric splitting method on the governing equations for a rigid spheroidal particle model with torques, drag and gravity. The method splits the operators into a vector field that is conservative and one that takes into account the forces of the fluid. Error analysis and numerical tests are performed on perturbed and stiff particle-fluid systems. For the perturbed case, the splitting method greatly improves the solution accuracy, when compared to a conventional multi-step method, and the global error behaves as $\mathcal{O}(\varepsilon h^2)$ for roughly equal computational cost. For stiff systems, we show that the splitting method retains stability in regimes where conventional methods blow up. In addition, we show through numerical experiments that the global order is reduced from $\mathcal{O}(h^2/\varepsilon)$ in the non-stiff regime to $\mathcal{O}(h)$ in the stiff regime.
\end{abstract}
\section{Introduction}%%%%%%%%%%%%%%%%%%%%%%%%%%%%%%%%%%%%%%%
%%%%%%%%%%%%%%%%%%%%%%%%%%%%%%%%%%%%%%%%%%%%%%%%%%%%%%%%%%%%%
%%%%%%%%%%%%%%%%%%%%%%%%%%%%%%%%%%%%%%%%%%%%%%%%%%%%%%%%%%%%%

Simulating the dynamics of particles in a fluid is of importance to many industrial applications such as paper making \cite{paper}, pharmaceutical processing \cite{biopolymer} and soot emission from combustion processes \cite{soot} as well as natural processes including the transportation of plankton in the sea \cite{plankton}, the formation of ice clouds \cite{iceclouds} and the dispersion of pollen in the atmosphere \cite{pollen}. With growing needs for larger models and longer simulation times, there is an increasing demand for effective numerical methods that minimise computational cost. Over the past 50 years, splitting methods have been used to model problems in molecular biology, physics and fluid dynamics, for example, and have been shown to supersede classical integration schemes in terms of both quantitative and qualitative accuracy \cite{splitting}. In this paper, we employ splitting methods on the axisymmetric rigid-body equations with Stokes viscous force, torque and gravity. Splitting methods are often used when the differential equation has geometric properties that should be preserved under disretisation, such as being Hamiltonian or divergence-free; or possessing a symmetry or a first integral. The idea behind splitting methods is to split the system into two or more simpler sub-systems and compute the numerical flow as the composition of the analytic flows of the subsystems at discrete time-steps. As these methods are purpose-built for the problem under study, they have the ability to mimic the qualitative behaviour of the continuous solution resulting in efficiency and stability improvements over standard, all-purpose integration techniques.\\

The particle-fluid system is modelled under the assumptions that the particle size is smaller than the smallest fluid length scale (e.g., the Kolomogrov scale) and that the particle shape can be approximated by a triaxial ellipsoid. Under the first assumption, the particle-Reynolds number is likely to be low and the fluid can be approximated by Stokes flow conditions where the dominant forces are drag, torque and gravity. We adopt the second assumption for numerous reasons. Due to the inherent complexity of fluid dynamics, ellipsoids are the only shape where the fluid forces are exactly known at leading order without making overly restrictive assumptions. For example, slender body theory can tell us the forces on the particle only but only if the particle is very long and thin \cite{slenderbody3, slenderbody2, slenderbody1} and perturbation theory can tell us the translational \cite{perturbation1} and rotational \cite{perturbation2} forces only for nearly spherical particles. Other than these two cases, the only shape where the forces are known at leading order are ellipsoids, which are modelled by Stokes viscous force, derived by Brenner \cite{brenner}, and torques, derived by Jeffery \cite{jeffery}. Such models have been adopted in studies such as \cite{mortensen,zhang,disks}. Additionally, modelling general non-spherical particles as axisymmetric spheroids, such as rigid rods \cite{zhang} or disks \cite{disks}, is a common leading order approximation, for example, ice-cloud particles are hexagonal plates and columns but are modelled as oblate and prolate spheroids \cite{iceclouds}. For a comprehensive review on particle modelling the reader is referred to \cite{voth}. In this paper we pay particular attention to two cases, one where the fluid forces are seen as a perturbation to an otherwise free rigid-body system and the second is a stiff system, where the fluid forces dominate the free rigid-body equations. \\

For non-spherical particles, the orientation couples with the translational dynamics and therefore greatly increases the model complexity. As a result, a system of 13 coupled ordinary differential equations (ODEs) need to be solved per time-step: three each for the position, velocity and angular momentum vectors and four for the rotation quaternion. A typical approach to solving these ODEs has been to integrate the system using Runge-Kutta methods and/or linear multi-step methods such as a second-order explicit Adams-Bashforth method \cite{mortensen,disks}. These methods, although straightforward to implement, present a number of drawbacks when calculating long-time numerical solutions to ODEs: (1) stability restrictions on the time-step $h$; (2) not time symmetric; and (3) limited ability to conserve properties specific to the underlying physics of the system. 

Such issues can only be overcome by enforcing small time-steps, thus increasing the total cost of the solution method, which limits the feasibility of large (e.g., $N>10^6$ particles) or long (e.g., $T\in [0,10^3]$ seconds) simulations \cite{voth}. Alternatively, one could approach the problem with a purpose-built algorithm, such as a splitting method, which takes advantage of particular properties of the vector field under study. Here, we show that when compared to a conventional two-step Adams-Bashforth method, the splitting method is both cheaper, more accurate and more robust thus allowing for larger time-steps to achieve the same accuracy. \\

The next section of the paper reviews relevant theory in particle modelling. We then introduce the numerical splitting method and present an error analysis. Section \ref{results} presents some numerical experiments and the last section is dedicated to conclusions.

%%%%%%%%%%%%%%%%%%%%%%%%%%%%%%%%%%%%%%%%%%%%%%%%%%%%%%%%%%%%%
%%%%%%%%%%%%%%%%%%%%%%%%%%%%%%%%%%%%%%%%%%%%%%%%%%%%%%%%%%%%%
\section{Governing equations}%%%%%%%%%%%%%%%%%%%%%%%%%%%%%%%%
%%%%%%%%%%%%%%%%%%%%%%%%%%%%%%%%%%%%%%%%%%%%%%%%%%%%%%%%%%%%%
%%%%%%%%%%%%%%%%%%%%%%%%%%%%%%%%%%%%%%%%%%%%%%%%%%%%%%%%%%%%%
  
\begin{figure}[h]
	\centering
	\includegraphics[width=0.7\linewidth]{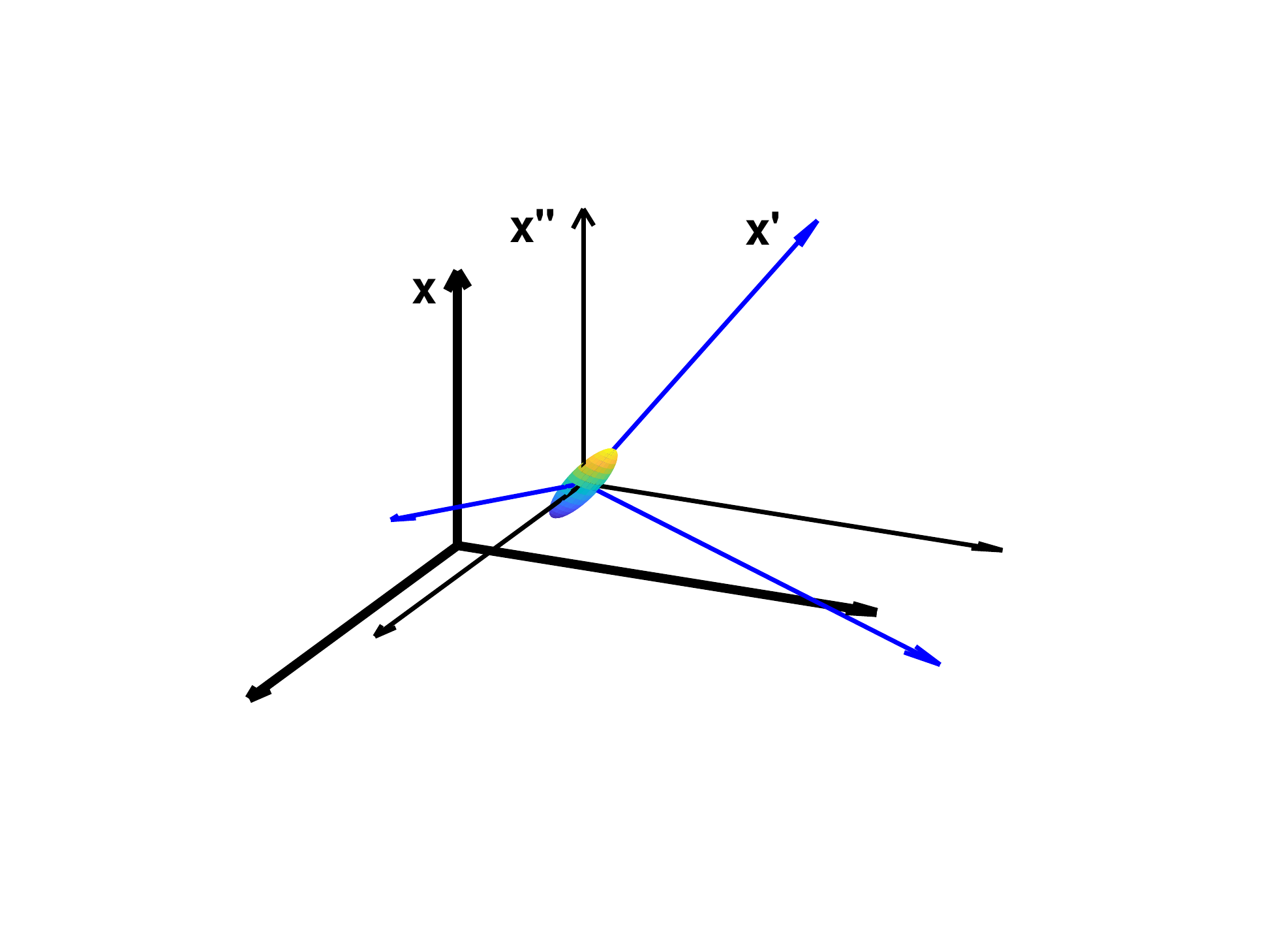}
	\caption{A prolate spheroid ($\lambda=3$) with coordinate lines of the inertial frame (thick black arrows), translating frame (thin black arrows) and the body frame (thin blue arrows). }
	\label{fig:reference_frames}
\end{figure}

  To describe the forces on the particle we first establish three reference frames. First, we define an \textit{inertial frame} by variables $\bx = (x,y,z)\trans$ that is an inertial coordinate system as shown in figure \ref{fig:reference_frames}. Secondly, we define a \textit{translating frame} by variables $\bx'' = (x'',y'',z'')\trans$ that is translating with the particle and has its origin co-located with the particles center of mass. Lastly, we introduce a \textit{body frame} denoted by variables $\bx' = (x',y',z')\trans$ that is translating and rotating with the particle. Henceforth, all primed and double primed variables are respectively defined in the body and translating frame and unprimed variables are defined in the inertial frame. \\
  
 Jeffery and Brenner derived forces for general rigid ellipsoids, which have three distinct semi-axis lengths; however, for simplicity we will focus on spheroids, which are axisymmetric. In the body frame, a spheroid is defined by
\begin{equation}\label{eqn:spheroid}
\frac{x'^2}{a^2}+\frac{y'^2}{a^2}+\frac{z'^2}{c^2} = 1,
\end{equation}
where $a$ and $c$ are the distinct semi-axis lengths. The particle shape is characterised by the dimensionless aspect ratio $\lambda = c/a>0$, which distinguishes between spherical ($\lambda = 1$), prolate ($\lambda > 1$) and oblate ($\lambda < 1$) particles (the latter two shapes are also called as rods and disks). The axisymmetric moment of inertia tensor for a spheroid in the body frame is
\begin{equation}
I' = ma^2\mathrm{diag}\left(\frac{(1+\lambda^2)}{5},\frac{(1+\lambda^2)}{5},\frac{2}{5}\right),
\end{equation}
where $m=\frac{4}{3}\pi \lambda a^3 \rho_p$ is the particle mass and $\rho_p$ is the particle density. \\

 A spheroid immersed in a fluid will experience forces on its surface that have magnitude governed by many parameters such as the particles density $\rho_p$, semi-major axis length $a$, aspect ratio $\lambda$, fluid density $\rho_f$, dynamic viscosity $\nu$ and fluid relaxation time $\tau_f$, which is defined in section \ref{ch:fluid}. Hence, it is a logical step to non-dimensionalise our equations by introducing a dimensionless Stokes number. The particle Stokes number is formally defined as the ratio of the particle and fluid relaxation times $St=\tau_p/\tau_f$. In this paper, we will adopt the definition 
\begin{equation}
St = \frac{D \lambda^2 a^2}{\nu \tau_f},
\end{equation}	
where $D = \frac{\rho_p}{\rho_f}$ is the particle-fluid density ratio. The Stokes number is a dimensionless measure of the relative importance of particle inertia, that is, as $St \rightarrow \infty $ the particle behaves as a free body and as $St \rightarrow 0$ the particle behaves as if itself were part of the fluid. Henceforth, all equations are presented in their non-dimensional form and all parameters have dimension equal to $1$.\\

The linear momentum, angular momentum and position can be described by the column vectors $\bp$, $\bL'$, $\bx\in\mathbb{R}^3$, and the orientation can be represented using Euler parameters \cite{CM}, i.e. a vector $q=(e_0,e_1,e_2,e_3)\in \mathbb{R}^4$ satisfying the constraint
\begin{equation}
1 = e_0^2+e_1^2+e_2^2+e_3^2,
\end{equation}
 that uniquely determines the orientation of the body frame relative to the axes of translating frame (and hence to the inertial frame subject to an additional translation). The Euler parameters were first used for particle modelling by Fan \cite{Fan} and are used in place of the conventional Euler angles to avoid singularities. Each $q$ uniquely determines a rotation matrix $Q\in SO(3)$ that transforms a vector in the body frame $\bx'$ to a vector in the translating frame $\bx''$ via 
\begin{equation}
	\bx''=Q\bx'.
\end{equation}
 There is a 2-to-1 correspondence between Euler parameters and $3\times 3$ rotation matrices given by the so called Euler-Rodriguez map $\mathcal{E}: q \mapsto Q$ \cite{splittingfrb}. Setting $\mathbf{e}=(e_1,e_2,e_3)$, the rotation matrix $\mathcal{E}(q)=Q$ is constructed via 
\begin{equation}
Q=\mathbb{1}+2e_0\hat{\mathbf{e}}+2\hat{\mathbf{e}}\hat{\mathbf{e}},
\end{equation}
 where $\mathbb{1}$ is the $3\times 3$ identity matrix and we have introduced the hat map $\widehat{\cdot}:\mathbb{R}^3 \rightarrow \mathfrak{so}(3)$ defined by
\begin{equation}
\left(
\begin{array}{c}
\omega_1\\
\omega_2\\
\omega_3\\
\end{array}
\right)
\mapsto 
\widehat{\bom} = \left( 
\begin{array}{ccc}
0 & -\omega_1 &  \omega_2 \\
\omega_1 & 0  & -\omega_3 \\
-\omega_2  & \omega_3 & 0   \\
\end{array} 
\right),
\end{equation} 
where $\mathfrak{so}(3)$ is the Lie algebra of $SO(3)$ containing $3\times3$ skew-symmetric matrices satisfying $ \mathbf{\bom}\times\mathbf{v} = \hat{\bom}\mathbf{v}$ for $\bom, \bv\in\mathbb{R}^3$. This gives the following expression for $Q$ explicitly in terms of the Euler parameters
\begin{equation}
Q = 
\left(
\begin{array}{ccc}
e_0^2+e_1^2-e_2^2-e_3^2 & 2(e_1e_2-e_0e_3) & 2(e_1e_3+e_0e_2) \\
2(e_1e_2+e_0e_3) & e_0^2-e_1^2+e_2^2-e_3^2 & 2(e_2e_3-e_0e_1) \\
2(e_1e_3-e_0e_2) & 2(e_2e_3+e_0e_1) & e_0^2-e_1^2-e_2^2+e_3^2 \\
\end{array}
\right).
\end{equation}

\subsection{Translational dynamics}
The Stokes viscous force, derived in \cite{brenner} and gravity force terms, are given in their non-dimensional form by
\begin{align}
\mathbf{F}_h =& \frac{3\lambda}{4St} QK'Q\trans(\mathbf{u}-\mathbf{v}), \label{eq:drag}\\
\mathbf{F}_g =& -m\mathbf{g},
\end{align}
 where $\bv$ is the inertial frame linear velocity, which is related to linear momentum via $\bp = m\bv$. Note that in our non-dimensional formalism we take $m=1$ to be a dimensionless constant; however, we will leave $m$ in our equations for consistency with the literature. The inertial frame fluid velocity vector $\bu = \bu(\bx,t)$ is taken at the location of the particle $\bx$ and the inertial frame gravity acceleration vector is $\mathbf{g}=(0,0,g)^T$ for some positive constant $g$ that is typically defined as $g=1-1/D$ to account for the buoyancy force. The body frame resistance tensor $K'$, derived by Oberbeck \cite{oberbeck}, is given by 

\begin{equation}
K'=16\pi\lambda~\mathrm{diag}\left( \frac{1}{\chi_0+\alpha_0},\frac{1}{\chi_0+\beta_0} ,\frac{1}{\chi_0+\lambda^2\gamma_0}\right)
\end{equation}

where the constants $\chi_0$, $\alpha_0$, $\beta_0$ and $\gamma_0$ were calculated for ellipsoidal particles by Siewert \cite{siewert} and are presented in table \ref{table:constants}. Note that the inertial frame resistance tensor $K$ is calculated from the similarity transformation $K=QK'Q\trans$.\\

\begin{table}[h]
\begin{center}
\renewcommand{\arraystretch}{1.8}
\begin{tabular}{c|ccc}
& $\lambda<1$ & $\lambda=1$ & $\lambda>1$ \\
\hline

$\chi_0$ 		& $\frac{\lambda^2(\pi-\kappa_0)}{\sqrt{1-\lambda^2}}$ & $2$ &  $\frac{-\kappa_0\lambda}{\sqrt{\lambda^2-1}}$\\

$\alpha_0 = \beta_0$ 	& $\frac{-\lambda \left(\kappa_0-\pi+2\lambda\sqrt{1-\lambda^2}\right)}{2(1-\lambda^2)^{3/2}}$ & $\frac{2}{3}$ & $\frac{\lambda^2}{\lambda^2-1}+\frac{\lambda\kappa_0}{2(\lambda^2-1)^{3/2}}$\\

$\gamma_0$	& $\frac{\left(\lambda(\kappa_0-\pi)+2\sqrt{1-\lambda^2}\right)}{(1-\lambda^2)^{3/2}}$ & $\frac{2}{3}$ & $\frac{-2}{\lambda^2-1}-\frac{\lambda\kappa_0}{(\lambda^2-1)^{3/2}}$\\

$\kappa_0$ 		& $2\arctan\left(\frac{\lambda}{\sqrt{1-\lambda^2}}\right)$ & $1$ & $\ln\left(\frac{\lambda-\sqrt{\lambda^2-1}}{\lambda+\sqrt{\lambda^2-1}}\right)$ \\
\end{tabular} 
\end{center}
\caption{The values for the constants $\chi_0$, $\alpha_0$, $\beta_0$ and $\gamma_0$ for $\lambda<1$, $\lambda=1$ and $\lambda>1$. }
\label{table:constants}
\end{table}

It will be convenient for the formulation of the methods to rewrite equation \eqref{eq:drag} as
\begin{equation}
\mathbf{F}_h = -A_1\bp+\mathbf{b}_1, \\
\end{equation}
where 
\begin{equation}\label{eq:A_1}
A_1=\frac{3\lambda}{4mSt}K \quad \mathrm{and} \quad
\mathbf{b}_1=mA_1\mathbf{u}(\bx,t).
\end{equation}
 Here, $\mathbf{b}_1$ is implicitly dependent on time through the fluid. This leads to the following ODE for momentum
\begin{equation}\label{eq:pdot}
\dot{\bp} = - A_1\bp+\mathbf{b}_1 - m\mathbf{g}.
\end{equation}
The inertial frame position vector $\bx$ is calculated by solving
\begin{equation}
	\dot{\mathbf{x}} = \bv. 
\end{equation}

\subsection{Rotational dynamics}

The rotational dynamics of an ellipsoidal particle are governed by the free rigid-body equations \cite{marsden} with torques $\mathbf{N}'=(N'_x,N'_y,N'_z)\trans$ that describe the rotational forces acting on an ellipsoid in creeping Stokes flow in the body frame \cite{jeffery}. These are presented in their non-dimensional form
\begin{align}
N'_x = & \frac{16\pi \lambda}{3(\beta_0+\lambda^2\gamma_0)}\left[(1-\lambda^2)S'_{yz}+(1+\lambda^2)(\Omega'_x-\omega'_x)\right],\label{eq:JT1} \\ 	
N'_y = & \frac{16\pi \lambda}{3(\alpha_0+\lambda^2\gamma_0)}\left[(\lambda^2-1)S'_{zx}+(1+\lambda^2)(\Omega'_y-\omega'_y)\right],\label{eq:JT2} \\ 
N'_z = & \frac{32\pi \lambda}{3(\alpha_0+\beta_0)}(\Omega'_z-\omega'_z),\label{eq:JT3}
\end{align}
where $\bom' = (\omega'_x,\omega'_y,\omega'_z)\trans$ is the body frame angular velocity, which is related to body frame angular momentum by $\bL' = I'\bom'$. The dimensionless body frame shear $\mathbf{S}'=(S'_{yz},S'_{zx},S'_{xy})\trans$ and fluid rotation $\boldsymbol{\Omega}'=(\Omega'_x,\Omega'_y,\Omega'_z)\trans$ terms are 
	\begin{equation}
S'_{ij} = \frac{1}{2}\left( \frac{\partial u'_i}{\partial x'_j} + \frac{\partial u'_j}{\partial x'_i} \right)\quad\mathrm{and}\quad\Omega'_i =  \frac{1}{2} (\nabla'\times\bu')_i.
\end{equation}
We write equations \eqref{eq:JT1}, \eqref{eq:JT2} and \eqref{eq:JT3} compactly as
\begin{equation}
\mathbf{N'} = -A'_2\bL'+\mathbf{b}'_2,
\end{equation}
where
\begin{equation}\label{eq:A_2}
A'_2 = \frac{12\lambda^2}{St}
\mathrm{diag}\left(\frac{(1+\lambda^2)}{(\beta_0+\lambda^2\gamma_0)},
\frac{(1+\lambda^2)}{(\alpha_0+\lambda^2\gamma_0)},
\frac{2}{(\alpha_0+\beta_0)}\right)I'^{-1},
\end{equation}
and 
\begin{equation}
\mathbf{b}'_2 = \frac{12\lambda^2}{St}
\mathrm{diag}\left(\frac{(1-\lambda^2)}{(\beta_0+\lambda^2\gamma_0)},
\frac{(\lambda^2-1)}{(\alpha_0+\lambda^2\gamma_0)},
0\right)\mathbf{S}'+A_2I'\boldsymbol{\Omega}'.
\end{equation}
Here, $\mathbf{b}'_2$ is implicitly dependent on time through the shear and rotation terms. The dimensionless equation governing the angular momentum of the particle in the body frame is therefore
\begin{equation}\label{eq:rotation}
\dot{\bL}'=\bL'\times\bom'-A'_2\bL'+\mathbf{b}'_2,
\end{equation}
where the cross-product term is the Poisson bracket for the free rigid-body \cite{marsden} that arises from the fact that $\bL'$ is represented in the (non-inertial) body frame. The rotation matrix $Q$ is calculated by solving the matrix ODE
\begin{equation}\label{eq:Qdot}
\dot{Q} = Q\widehat{\bom}', 
\end{equation}
which arises from the quaternion formulation for the rigid-body, see \cite{splittingfrb} for details. When designing a splitting method, it is notationally convenient to express the ODEs as vector equations. To do so we will denote $\bq_i$ to be the $i$th column of $Q\trans$, then
\begin{equation}\label{eq:qidot}
	\dot{\mathbf{q}_i}=-\widehat{\bom}'\mathbf{q}_i \quad \mathrm{for} \quad i=1,2,3 
\end{equation}
which represents three vector equations. It is important to stress, that to ensure that the orthogonality of $Q$ is preserved, it is equation \eqref{eq:Qdot} that is being solved during the implementation of the splitting method and not equation \eqref{eq:qidot}.
 \subsection{Fluid field}\label{ch:fluid}
 This paper is only concerned with the performance of numerical methods in calculating solutions to particle dynamics, so as to measure this in isolation of the costs associated with discrete fluid field interpolation, an analytic fluid field that is known everywhere in time and space is used. The inertial frame fluid velocity vector $\bu=(u,v,w)\trans$ is modelled by an analytic solution to the Navier-Stokes equations derived by Ethier and Steinman \cite{etheir-steinman} 
 \begin{align}
 u =& -\alpha_f[\ee^{\alpha_f x}\sin(\alpha_f y\pm \beta_f z)+\ee^{\alpha_f z}\cos(\alpha_f x\pm \beta_f y)]\ee^{-\beta_f ^2t},\label{fluidx}\\
 v =& -\alpha_f[\ee^{\alpha_f y}\sin(\alpha_f z\pm \beta_f x)+\ee^{\alpha_f x}\cos(\alpha_f y\pm \beta_f z)]\ee^{-\beta_f ^2t},\label{fluidy}\\
 w =& -\alpha_f[\ee^{\alpha_f z}\sin(\alpha_f x\pm \beta_f y)+\ee^{\alpha_f y}\cos(\alpha_f z\pm \beta_f x)]\ee^{-\beta_f ^2t},\label{fluidz}
 \end{align} 
 for positive constants $\alpha_f$ and $\beta_f$. The fluid model has time scale $\tau_f = \beta_f^{-2}$ and is chosen as it has non-zero, non-trivial velocities that depend on every direction in each component of $\bu$ and its Jacobian $\nabla\bu$, and is derived from the full Navier-Stokes equation (i.e., without neglecting the convective, diffusive, unsteady or pressure terms). We assert that this fluid field provides a reasonable test of the solution methods in a non-trivial fluid and insights into their performance when the flow is transitioned to a realistic field, for example in \cite{mortensen,disks,zhang}. In addition, we will conduct long-time experiments on an oscillating shear flow field defined by $\bu_S=(0,0,x\cos(2\pi t)/\tau_f)\trans$.\\

%%%%%%%%%%%%%%%%%%%%%%%%%%%%%%%%%%%%%%%%%%%%%%%%%%%%%%%%%%%%%
%%%%%%%%%%%%%%%%%%%%%%%%%%%%%%%%%%%%%%%%%%%%%%%%%%%%%%%%%%%%%
\section{Numerical methods}%%%%%%%%%%%%%%%%%%%%%%%%%%%%%%%%%%%%%%%%%%%%
%%%%%%%%%%%%%%%%%%%%%%%%%%%%%%%%%%%%%%%%%%%%%%%%%%%%%%%%%%%%%
%%%%%%%%%%%%%%%%%%%%%%%%%%%%%%%%%%%%%%%%%%%%%%%%%%%%%%%%%%%%%

\subsection{Splitting}
Splitting methods can be used when an ODE can be expressed as the sum of two or more operators, 
\begin{equation}\label{eq:split ODE}
	\dot{\by}(t) = f(\by) = f_1(\by) + f_2(\by),
\end{equation}
where $\by\in \mathbb{R}^n$ and $f_1$, $f_2:\mathbb{R}^n\rightarrow\mathbb{R}^n$. Ideally, the splitting is chosen in such a way that the flows\footnote{We denote by $\varphi_h$ the flow operator such that $\by(h) = \varphi_h(\by_0)$ is the solution of the ODE at time $t=h$ with initial conditions $\by_0$ at $t=0$.} $\varphi^{[1]}_h$ and $\varphi^{[2]}_h$ of the systems $\dot{\by}(t) = f_1(\by)$ and $\dot{\by}(t) = f_2(\by)$ can be computed exactly. In this case, numerical approximations can be generated by
\begin{equation}\label{eq:LT}
\Phi_h = \varphi^{[1]}_h\circ\varphi^{[2]}_h ,\quad\mathrm{or}\quad \Phi_h^* = \varphi^{[2]}_h\circ\varphi^{[1]}_h,
\end{equation}
which are known as Lie-Trotter splittings \cite{trotter} and are each others adjoints. Taylor expansion shows that the method is first-order. Another numerical method can be generated by 
\begin{equation}\label{eq:strang}
\Phi_h^{[S]}=	\varphi^{[1]}_{h/2}\circ\varphi^{[2]}_h\circ\varphi^{[1]}_{h/2},
\end{equation}
which is the Strang splitting method \cite{strang}. Note that this can be written as the composition of the above Lie-Trotter methods with half time-steps $\Phi^{[S]}_h=\Phi_{h/2}\circ\Phi^*_{h/2}$, hence the method is of second-order and is symmetric \cite[pg. 45]{GNI}. Similarly, $\Phi^{[S]}_h=\Phi^*_{h/2}\circ\Phi_{h/2}$ is also a second-order symmetric method. Symmetric methods of arbitrarily high order can be generated by composition of the above methods, however, we refer the reader to \cite{yoshida, suzuki} for a more complete description of high-order splitting methods. For a full review of splitting theory, we refer the reader to \cite{splitting}. 
 
\subsection{System of differential equations}

 Let $\by(t)=(\bp\trans,\bL{'}\trans,\bq_1\trans,\bq_2\trans,\bq_3\trans,\bx\trans)\trans\in\mathbb{R}^{18}$ be the solution to the ODE in the form of equation \eqref{eq:split ODE}. The particles dynamics is governed by the following system of first-order coupled ODEs
\begin{equation}\label{eq:system}
  \left.\begin{aligned}
\dot{\bp} &= -A_1 \bp + \mathbf{b}_1 - m\mathbf{g}, \\
\dot{\bL}' &= \bL'\times\bom'-A'_2 \bL'+\mathbf{b}'_2, \\
\dot{\mathbf{q}_i}&=-\widehat{\bom}'\mathbf{q}_i, \quad \mathrm{for} \quad i=1,2,3\\
\dot{\mathbf{x}} &= \bv, \\ 
\end{aligned} \right\} f(\by)
\end{equation}

where the RHS of the equations in \eqref{eq:system} arises due to the vector field $f(\by)$. The kinetic and potential energies $K$ and $U$, and Hamiltonian $H$ are given by 
\begin{align}
K(\mathbf{y}) = & \frac{1}{2}\bp\trans m^{-1} \bp +\frac{1}{2} \bL{'}\trans I'^{-1} \bL', \\	
U(\mathbf{y}) = & \frac{1}{2}\left(\mathbf{q}_1\trans \mathbf{q}_1 + \mathbf{q}_2\trans \mathbf{q}_2 + \mathbf{q}_3\trans \mathbf{q}_3\right) +m\bx\trans\mathbf{g},\\
{H}(\mathbf{y}) = & K(\mathbf{y})+U(\mathbf{y}),\label{eq:energy}
\end{align}
where $\left.\sum_{i=1}^{3}\bq\trans_i\bq_i=3\right.$ is a constant. The gradient of the Hamiltonian is 
\begin{equation}
\nabla{H}(\by) = \left( \bv\trans,{\bom'}\trans,\bq_1\trans,\bq_2\trans,\bq_3\trans,m\mathbf{g}\trans \right)\trans,
\end{equation}
and is related to the solution vector $\by$ by the following non-injective mapping
\begin{equation}
	\nabla H = M \by + \mathbf{g}_1, 
\end{equation}
 where the matrix $M := \mathrm{diag}(m^{-1}\mathbb{1},I^{-1},\mathbb{1},\mathbb{1},\mathbb{1},\mathbb{0})\in\mathbb{R}^{18\times 18}$ is diagonal and singular and $\mathbf{g}_1 = (0,\cdots,0,m\mathbf{g}\trans)\trans \in \mathbb{R}^{18}$. Now $\dot{\by}$ can be written as 
\begin{equation}\label{S}
\dot{\by} = f(\by) =  S\nabla{H}-A\by+\mathbf{b},
\end{equation}
where $S\in\mathbb{R}^{18\times18}$ is a skew-symmetric matrix given by
\begin{equation}
S = \left(\begin{array}{cccccc}
\mathbb{0} &\mathbb{0} &\mathbb{0} &\mathbb{0} &\mathbb{0} &-\mathbb{1} \\
\mathbb{0} &\widehat{\bL}' &\mathbb{0} &\mathbb{0} &\mathbb{0} &\mathbb{0} \\
\mathbb{0} &\mathbb{0} &-\widehat{\bom}' &\mathbb{0} &\mathbb{0} &\mathbb{0} \\
\mathbb{0} &\mathbb{0} &\mathbb{0} &-\widehat{\bom}' &\mathbb{0} &\mathbb{0} \\
\mathbb{0} &\mathbb{0} &\mathbb{0} &\mathbb{0} &-\widehat{\bom}' &\mathbb{0} \\
\mathbb{1} &\mathbb{0} &\mathbb{0} &\mathbb{0} &\mathbb{0} &\mathbb{0} \\
\end{array}\right),\\
\end{equation}
 $A\in\mathbb{R}^{18\times18}$ is a diagonal matrix given by
\begin{equation}
	A = \mathrm{diag}(A_1,A'_2,\mathbb{0},\dots,\mathbb{0}),
\end{equation}
 $\mathbf{b}\in\mathbb{R}^{18}$ is a vector given by 
\begin{equation}
\mathbf{b} = (\mathbf{b}_1\trans,\mathbf{b}{'}_2\trans,0,\dots,0)\trans\in\mathbb{R}^{18},
\end{equation}
% $B\in\mathbb{R}^{18\times18}$ is a matrix given by 
%\begin{equation}
%S = \left(\begin{array}{cccccc}
%\mathbb{0} &\mathbb{0} &\mathbb{1}\otimes\mathbf{b}_1(t) &\mathbb{0} \\
%\mathbb{0} &\mathbb{0} &\mathbb{1}\otimes\mathbf{b}_2(t) &\mathbb{0} \\
%\vdots &~ &\dots &\vdots \\
%\mathbb{0} &\dots &\dots &\mathbb{0} \\
%\end{array}\right),\\
%\end{equation}
and $\mathbb{0}\in\mathbb{R}^{3\times3}$ is the zero matrix. Note from equations \eqref{eq:A_1} and \eqref{eq:A_2} that matrices $A_1$ and $A_2'$ are positive definite, hence $A$ is positive semi-definite and therefore represents a linear dissipation. Additionally, vectors $\mathbf{b}_1$ and $\mathbf{b}_2'$ represent the forces of the fluid on the particle, hence $\mathbf{b}$ is a non-conservative force term. As the energy of such a system is necessarily non-constant, we can calculate the exact energy dissipation by taking the time derivative of the Hamiltonian
\begin{equation}\label{eq:dissipation}
		\dot{{H}} = \nabla{H}\trans\dot{\mathbf{y}}
	=\nabla{H}\trans(-A\by+\mathbf{b}),
\end{equation}
where we have used the fact that $\nabla{H}\trans S \nabla{H} = 0$ for skew-symmetric matrix $S$. 
With the forethought that we would like a dissipation-preserving splitting scheme, we split $f(\by)$ into the following two sub-systems
\begin{align}
\dot{\boldsymbol{\by}} = & f_1(\by) = S\nabla{H},\label{eq:f1}  \\
\dot{\boldsymbol{\by}} = & f_2(\by) = -A\by+\mathbf{b}.\label{eq:f2} 
\end{align}
The first system is Hamiltonian and hence $\dot{H}^{[1]}=0$ while the second system dissipates energy according to $\dot{H}^{[2]}=\nabla{H}\trans f_2(\by)=\nabla{H}\trans(-A\by+\mathbf{b})$. Hence, the numerical flow given by equation \eqref{eq:strang} preserves, up to the order of the method, the energy dissipation of the continuous system given by equation \eqref{eq:dissipation}. Equations \eqref{eq:f1} and \eqref{eq:f2} correspond to the following systems of ODEs
\begin{equation}\label{eq:system1}
\begin{array}{ccc}
\left.\begin{aligned}
\dot{\bp} &= - \mathbf{g}\\
\dot{\bL}' &= -\widehat{\bom}'\bL' \\
\dot{\mathbf{q}_i}&=-\widehat{\bom}'\mathbf{q}_i \\
\dot{\mathbf{x}} &= \bv \\ 
(\dot{t} &= 1) \\ 
\end{aligned} \right\}f_1(\by) 
& 
\mathrm{and} \quad 
&
\left.\begin{aligned}
\dot{\bp} &= -A_1 \bp +\mathbf{b}_1 \\
\dot{\bL}' &= -A'_2 \bL'+\mathbf{b}'_2 \\
\dot{\mathbf{q}_i}&=\mathbf{0} \\
\dot{\mathbf{x}} &= \mathbf{0} \\
(\dot{t} &= 0) \\  
\end{aligned} \right\}f_2(\by) ,
\end{array}
\end{equation}
where $f_1(\by)$ represents a free rigid-body vector field with gravity, while $f_2(\by)$ represents a purely energy dissipative (exponential decaying) vector field with a non-conservative force that leaves $Q$ and $\bx$ constant. Note that we freeze the flow of time in the second system to remove any implicit time dependence that $\mathbf{b}_1$ and $\mathbf{b}'_2$ may have through the fluid vector field.
 
%The system is made autonomous the usual way: by treating time as a dependent variable, propagating it only in $f_1(\by)$ and freezing it in $f_2(\by)$. This removes the time dependence of $\mathbf{b}$. 

\subsubsection{Solutions to $f_1(\by)$}
The original system of ODEs is split such that the resulting sub-systems have solutions that can be computed analytically. The first system is solved using the well known solutions for axisymmetric rigid bodies \cite[chapt. VII.5]{GNI}. Note that this method can be generalised to triaxial ellipsoids, see for example \cite{marsden,splittingfrb,frb}. First, the angular velocity $\bom'$ is solved by
\begin{equation}
\bom'(h) = R'_z(\mu h)\bom'_0,
\end{equation}
where $\mu =\omega'_z(0)\frac{I'_x-I'_z}{I'_x}$ and $R'_z(\mu h)$ is a planar rotation of angle $\mu h$ about the $z'$ axis of the body frame
\begin{equation}
R'_z(\mu h) = 
\left(\begin{array}{ccc}
\cos(\mu h) & \sin(\mu h) & 0 \\
-\sin(\mu h) & \cos(\mu h) & 0 \\
0 & 0 & 1 \\
\end{array}\right).
\end{equation}
This immediately yields the angular momentum
\begin{equation}
\bL'(h)=I'\bom'(h).
\end{equation}
 Next, setting $w(h)=(0,\bom'(h)\trans)\trans\in\mathbb{R}^4$, the rotation matrix $Q=\mathcal{E}(q)$ is solved by computing the quaternion
\begin{equation}
q(h)=q_0\cdot\mathrm{exp_q}\left(\frac{h}{2}w(h/2)\right),
\end{equation} 
where $\mathrm{exp_q}$ is the quaternion exponential and the $\cdot$ represents multiplication of two quaternions (see \cite{splittingfrb} for details). Here, $w(h/2)$ is evaluated at a half time-step to maintain symmetry. The linear momentum $\bp$ is solved by 
\begin{equation}
\bp(h)=-m\mathbf{g}h+\bp_0   ,
\end{equation}
and the position $\bx$ is calculated by integrating the velocity
\begin{equation}
\bx(h)=-\frac{1}{2}\mathbf{g}h^2 + \frac{1}{m}\bp_0 h+\bx_0. 
\end{equation}
These solutions to $\bp$, $\bL'$, $Q$ and $\bx$ at time $h$ are represented by the flow map $\varphi^{[1]}_h$ in equation \eqref{eq:strang}.
 \subsubsection{Solutions to $f_2(\by)$}
The $\bL'$ and $\bp$ equations in $f_2(\by)$ of equation \eqref{eq:system1} are solved using the variation of constants formula
\begin{align}
\bp(h) &= \exp\left(A_1h\right)\left(\bp_0+A_1^{-1}\mathbf{b}_1\right)-A_1^{-1}\mathbf{b}_1,\\
\bm'(h) &= \exp\left(A'_2h\right)\left(\bm'_0+A{'}_2^{-1}\mathbf{b}'_2\right)-A{'}_2^{-1}\mathbf{b}'_2.
\end{align}
Where vectors $\mathbf{b}_1$ and $\mathbf{b}'_2$ are constant in this system as we have enforced $\dot{t}=0$. Additionally, the rotation matrix $Q$ and the position vector $\bx$ are also kept constant in this system. These solutions at time $h$ are represented by the flow map $\varphi^{[2]}_h$ in equation \eqref{eq:strang}.

\section{Error Analysis}
The dissipative system $f_2(\by)$ that represents the fluid forces is inversely proportional to the Stokes number $St$ which can be taken to be small ($St<<1$) or large ($St>>1$), depending on the application. In addition, the choice of $\lambda$ can greatly effect the magnitude of matrix $A$ and vector $\mathbf{b}$. In fact it can be shown that  $||A||\leq c_1 \lambda^4/St$ for $\lambda>1$ and $||A||\leq c_2\sqrt{\lambda}/St$ for $\lambda<1$ (see table \ref{table:constants}) and for some positive constants $c_1$ and $c_2$. This leads us to consider at least two main cases: one where $f_2(\by) = \varepsilon \tilde{f_2}(\by)$ is a perturbation and another where $f_2(\by) = \frac{1}{\varepsilon}\tilde{f_2}(\by)$ is a stiff term for $0<\varepsilon<<1$. For the remainder of this section we will set $\mathbf{b}=\mathbf{0}$  (i.e. that $f_2$ consists only of a linearly dissipative term) and assume that gravity is negligible such that $\nabla H \approx M\by$. We will use backward error analysis to study the error in the non-stiff case, and we will illustrate the behaviour of the error in the stiff case by numerical tests. We will let $\gamma^i$ represent the eigenvalues of the dissipation matrix $A$, of which six are non-zero and are the diagonal elements of matrices $A_1'=Q\trans A_1 Q$ and $A_2'$, given in equations \eqref{eq:A_1} and \eqref{eq:A_2} respectively.\\

The local error for the energy $H$ is given by the scalar
\begin{equation}
	\delta_H(\by_0) = H(\by(h)) - H(\by_1).
\end{equation}
If gravity is negligible, the particles energy is only kinetic, hence $ H = \frac{1}{2}\by\trans M\by = \frac{1}{2}\nabla H \trans\by$. Using the fact that the numerical approximation $\by_1$ differs from the exact solution $\by(h)$ by the local error $\by_1 = \by(h)+\boldsymbol{\delta}(\by_0)$, it follows that the local energy error reduces to 
\begin{equation}
	\delta_H(\by_0) = -\nabla H\trans 
	\bd(\by_0) + \mathcal{O}(||\bd||^2).\label{eq:Hle} \\
%\delta^{[\dot{H}]}(\by_0) =& -2\nabla H \trans A \bd(\by_0) + \mathcal{O}(\bd^2).\label{eq:Hdotle}
\end{equation}
The next section will be dedicated to calculating the local solution error $\bd(\by_0)$ and local energy error $\delta^{[H]}(\by_0)$ for the numerical method for the perturbed case. For the stiff case we will explore the global error using numerical experiments. 
\subsection{Non-stiff case} 

Here, we will look at a modified vector field that coincides exactly with the flow of the numerical method and compare this to the exact vector field. For $\gamma^i<<1$, we can write the ODE as $\dot{\by} = f_1(\by) + f_2(\by)= f_1(\by) + \varepsilon \tilde{f_2}(\by)$. Here, we have introduced the scaled variables, denoted by the tilde, in our case $\varepsilon\tilde{f}_2 = \varepsilon \tilde{A} \by$. For arguments sake, we will analyse the error for the Lie-Trotter splitting as the results are more concise and analogous to the Strang splitting method. The numerical flow corresponding to the Lie-Trotter operator is 
%\begin{equation}
%	\begin{array}{lll}
%	\mathrm{Lie-Trotter:} & \Phi_h^{[LT]}(\by_0) = \exp(f_1h)\exp( f_2h), & (\mathrm{Equation~\eqref{eq:LT}} )\\
%	\mathrm{Strang:} &  \Phi^{[S]}_h(\by_0) = \exp( f_2\frac{h}{2})\exp(f_1h)\exp( f_2\frac{h}{2}). & (\mathrm{Equation~ \eqref{eq:strang}})
%	\end{array}
%\end{equation}
\begin{equation}
 \Phi_h^{[LT]}(\by_0) = \varphi^{[1]}_h\circ\varphi^{[2]}_h(\by_0),
\end{equation}
 The local error can be determined by taking the difference between the exact and numerical flow over one time-step starting from the initial conditions $\by(0) = \by_0$. It follows that the local error for the Lie-Trotter method is 
\begin{equation}\label{eq:LTle}
 \bd^{[LT]}(\by_0) =\varphi_h(\by_0)-\Phi^{[LT]}_h(\by_0)= \frac{h^2}{2}[f_1,f_2]\mathbf{y}_0+\mathcal{O}(h^3),\\
\end{equation}
%and
%\begin{align}\label{eq:strangle}
%	 \bd^{[S]}(\by_0) =& \varphi_h(\by_0)-\Phi^{[S]}_h(\by_0)\nonumber\\
%	=& h^3 (\frac{1}{12}[f_1,[f_1,f_2]]-\frac{1}{24}[f_2,[f_2,f_1]])\mathbf{y}_0+\mathcal{O}(h^4),
%\end{align}
 where we have Taylor expanded the flows and use the bilinear Lie bracket of vector fields \cite[chapt. IV]{GNI}, which expressed in coordinates is given by
\begin{equation}
[f_1,f_2] = \sum_{i,j=1}^{n}\left( f_1^i\partial_i f_2^j - f_2^i\partial_i f_1^j \right)\partial_j,
\end{equation}
where $f^j_1$ is the $j$th element of $f_1$ and $\partial_i=\partial/\partial y^i$. Inserting equations \eqref{eq:f1} and \eqref{eq:f2} into \eqref{eq:LTle} we can write the local error explicitly  
\begin{equation}\label{LTerror}
\bd^{[LT]}(\by) =   \frac{ \varepsilon h^2}{2}(S\nabla^2H\tilde{A}\by - \tilde{A}S\nh - \nabla \tilde{A}(S\nh,y,\cdot) -  \nabla S(\tilde{A}\by,\nabla H,\cdot)) + \mathcal{O}(h^3),
\end{equation}
where the tri-linear tensor $\nabla S$ is calculated by taking the gradient of $S$ and satisfies the skew-symmetric relationship $\nabla S(\mathbf{u},\mathbf{v},\mathbf{w}) = - \nabla S(\mathbf{u},\mathbf{w},\mathbf{v})$ in it's last two components for vectors $\mathbf{u}$,$\mathbf{v}$,$\mathbf{w}\in\mathbb{R}^{18}$, hence $\nabla S(\tilde{A}\by,\nabla H,\cdot)$ is interpreted as a column vector. If we insert equation \eqref{LTerror} into equation \eqref{eq:Hle} we can compute the local energy error 
\begin{align}
\delta^{[LT]}_H(\by) =&   \frac{\varepsilon h^2}{2}(\nh\trans S\nabla^2H\tilde{A}\by - \nh\trans \tilde{A}S\nh - \nabla \tilde{A}(S\nh,y,\nh ) -  \nabla S(\tilde{A}\by,\nabla H,\nabla H)) + \mathcal{O}(h^3),\nonumber\\
=& \frac{\varepsilon h^2}{2}(\nh\trans S\nabla^2H\tilde{A}\by - \nh\trans \tilde{A}S\nh - \nabla \tilde{A}(S\nh,y,\nh )) + \mathcal{O}(\varepsilon h^3),
\end{align}
where, we have used the fact that $\nabla S(\tilde{A}\by,\nabla H,\nabla H)$ vanishes due to the skew-symmetry of its last two components.\\

To compute the global error of the Lie-Trotter method, we first assume that both vector fields $f_1$ and $f_2$ are one-sided Lipschitz with $L_1$ and $L_2$ as their respective one-sided Lipschitz constants. We will also use a result of \cite[pg. 37]{SODE1}, which states that for a first-order ODE that has two solutions $\by_1(h)$ and $\by_2(h)$, their difference is bounded by the inequality
\begin{equation}\label{odebound}
||\by_1(h) - \by_2(h)|| \leq \ee^{Lh}||\by_1(0)-\by_2(0)||,
\end{equation}
for one-sided Lipschitz constant $L$. The global error at time $t=t_{n+1}$ is
\begin{align}
	\be^{[LT]}_{n+1} =& \by(t_{n+1}) - \by_{n+1} \nonumber\\
	=& \bphi^{[LT]}_h(\by(t_n))+\bd^{[LT]}(\by(t_n)) - \bphi^{[LT]}_h(\by_n),
\end{align} 
which is computed by decomposing $\bphi^{[LT]}_h = \varphi^{[1]}_h\circ\varphi^{[2]}_h$ into its flow operators as follows 
\begin{align}
||\bphi^{[LT]}_h(\by(t_n)) - \bphi^{[LT]}_h(\by_n)|| &= ||\varphi^{[1]}_h\circ\varphi^{[2]}_h
\by(t_n)-\varphi^{[1]}_h\circ\varphi^{[2]}_h\by_n||\nonumber\\
&\leq \ee^{L_1 h} || \varphi^{[2]}_h \by(t_n) - \varphi^{[2]}_h \by_n ||,\nonumber\\
&\leq \ee^{(L_1+L_2) h} ||e^{[LT]}_n||, 
\end{align}
where we have used inequality \eqref{odebound} twice. If we then assume that the local error is bounded by $\varepsilon h^2 d \geq ||\bd^{[LT]}(y(t))||$, $\forall t\in[0,T]$ for some constant $d$ and for sufficiently small $h$, then 
\begin{equation}
	||\be^{[LT]}_{n+1}|| \leq \ee^{(L_1+L_2)h} ||\be^{[LT]}_n|| + ||\varepsilon h^2 d||, 
\end{equation}
this implies that the global error is bounded as follows
\begin{equation}
	||\be^{[LT]}_{n+1}|| \leq \varepsilon h^2||d\sum_{i=0}^{n}\left(\ee^{h(L_1+L_2)}\right)^i||,
\end{equation}
  where $n=T/h$. Taylor expanding the exponential shows the sum is $\mathcal{O}(1/h)$. We can therefore conclude that the global error magnitude is $||\be^{[LT]}_{n+1}||\sim\mathcal{O}(\varepsilon h)$.\\ 

The same argument of calculating the local error can be applied to the Strang method and although straightforward, involves the computation of nested commutator brackets. The results, however, are analogous and the local error $\bd^{[S]}(\by)$ is presented in appendix \ref{app:strang}. We find that the local error for the Strang splitting is $||\bd^{[S]}(\by)||\sim\mathcal{O}(\varepsilon h^3)$ at leading order and terms proportional to $\mathcal{O}(\varepsilon^2h^3)$ can be ignored for $\varepsilon<h$. It then follows that the global error of the Strang method is $|\be^{[S]}_{n+1}|\sim\mathcal{O}(\varepsilon h^2)$.\\

 For conventional one-step or multistep methods, such as the Adams-Bashforth two-step method, the perturbed and non-perturbed parts of the vector field are treated together, which means that the method does not see any error advantages due to the small parameter $\varepsilon$. As such, the global error is independent of $\varepsilon$. Using Taylor series it can be shown \cite[chapt. III]{SODE1} that the global error of the Adams-Bashforth two-step method is 
 \begin{equation}
	||\be_{n+1}^{[AB]}|| \sim \frac{5h^2}{12}||f''(\by_n)|| + \mathcal{O}(h^3),
\end{equation}
which is $\mathcal{O}(h^2)$ as opposed to the Strang splitting method which is $\mathcal{O}(\varepsilon h^2)$.

\subsection{Stiff case}
In this section we will examine the error of the splitting method when the vector field $f_2(\by)$ is stiff (i.e., when $\gamma^i>>1$). The differential equation can then be represented by $\dot{\by} = f_1(\by) + \frac{1}{\varepsilon} \tilde{f_2}(\by)$. A classical error analysis can be used in the non-stiff regime $h<\varepsilon$, and this shows that the global error behaves according to $\mathcal{O}(h^2/\varepsilon)$. However, in practise, one would like to use a step size $h>\varepsilon$ and in this situation, the flow operator $\varphi^{[2]}_h$ becomes somewhat more difficult to analyse because $||\frac{1}{\varepsilon} \tilde{f_2}(\by)||\geq 1$ and we cannot expand the flow of $f_2$ in its Taylor series, hence the classical error analysis fails when taking a Taylor expansion about the initial point of this flow operator. Many authors have studied the local error of various first- and second-order splitting methods in this situation using other means, such as singular perturbation theory \cite{kozlov,sportisse} or Lie series \cite{kozlov}. In these studies, it is shown that in the regime $h<\varepsilon$ the local error behaves according to the classical theory; however, for $h>\varepsilon$ different order reduction phenomena are observed depending on the splitting operator. These studies were performed in the context of designing robust splitting methods that use step size control based on local error estimates; however, we are primarily interested in the behaviour of the global error. There has been somewhat less research into how the global error behaves in the stiff case or how the order reduction in the local error evolves when measuring the global error of ODEs. Here, we present numerical experiments relating the local and global error to the step size $h$ and stiffness parameter $\varepsilon$. The results are presented in the next section. \\

\section{Numerical Results} \label{results}%%%%%%%%%%%%%%%%%%
%%%%%%%%%%%%%%%%%%%%%%%%%%%%%%%%%%%%%%%%%%%%%%%%%%%%%%%%%%%%%
%%%%%%%%%%%%%%%%%%%%%%%%%%%%%%%%%%%%%%%%%%%%%%%%%%%%%%%%%%%%%
Numerical tests were performed for a perturbed and stiff fluid-particle system in the 3D flow field described by equations \eqref{fluidx}, \eqref{fluidy} and \eqref{fluidz}. Numerical solutions are calculated using the second-order splitting method (SP2) and the second-order Adams-Bashforth two-step method (AB2) for comparison. The perturbed system uses the values $\lambda=0.1$, $St=100$, and the maximum eigenvalue of the dissipation matrix $A$ is $\gamma_{max}\approx0.0806$. The stiff system uses the values $\lambda=10$, $St=1$, and $\gamma_{max}\approx24,062$. Both systems use gravity and 3D fluid terms of $g = 0.99$, $\alpha_f=2\pi$ and $\beta_f = \pi$. The initial conditions for both experiments are $\bp_0 = (1,1,1)\trans$, $\bL_0 = (1,1,1)\trans$, $\bx_0 = (0,0,0)\trans$ and $q_0 = (1/\sqrt{2},0,1/\sqrt{2},0)\trans$ is the initial rotation quaternion. The error presented in the following figures is 
\begin{equation}
%	e^{[\bp]}_n = \frac{|\bp(t_n)-\bp_n|}{|\bp(t_n)|}
\mathrm{error} = \frac{||\by_n-\by(t_n)||}{||\by(t_n)||},
\end{equation}
 where $\by(t_n)$ is a reference solution calculated using the classical Runge-Kutta fourth-order method with a comparatively small time-step (e.g., $h=2^{-14}$). \\

%Figure \ref{fig:paths} presents examples of particle paths over the time interval $T=[0,1]$ for the perturbed and stiff systems. The fluid velocity vectors are shown in blue. It is seen that, in the stiff case, the particle trajectory is aligned with direction of the fluid vectors after the first snapshot, while in the perturbed case, the fluid has less of an influence over the particles trajectory. \\

\begin{figure}
	\begin{subfigure}{0.45\textwidth}
		\centering
		\includegraphics[width=\textwidth]{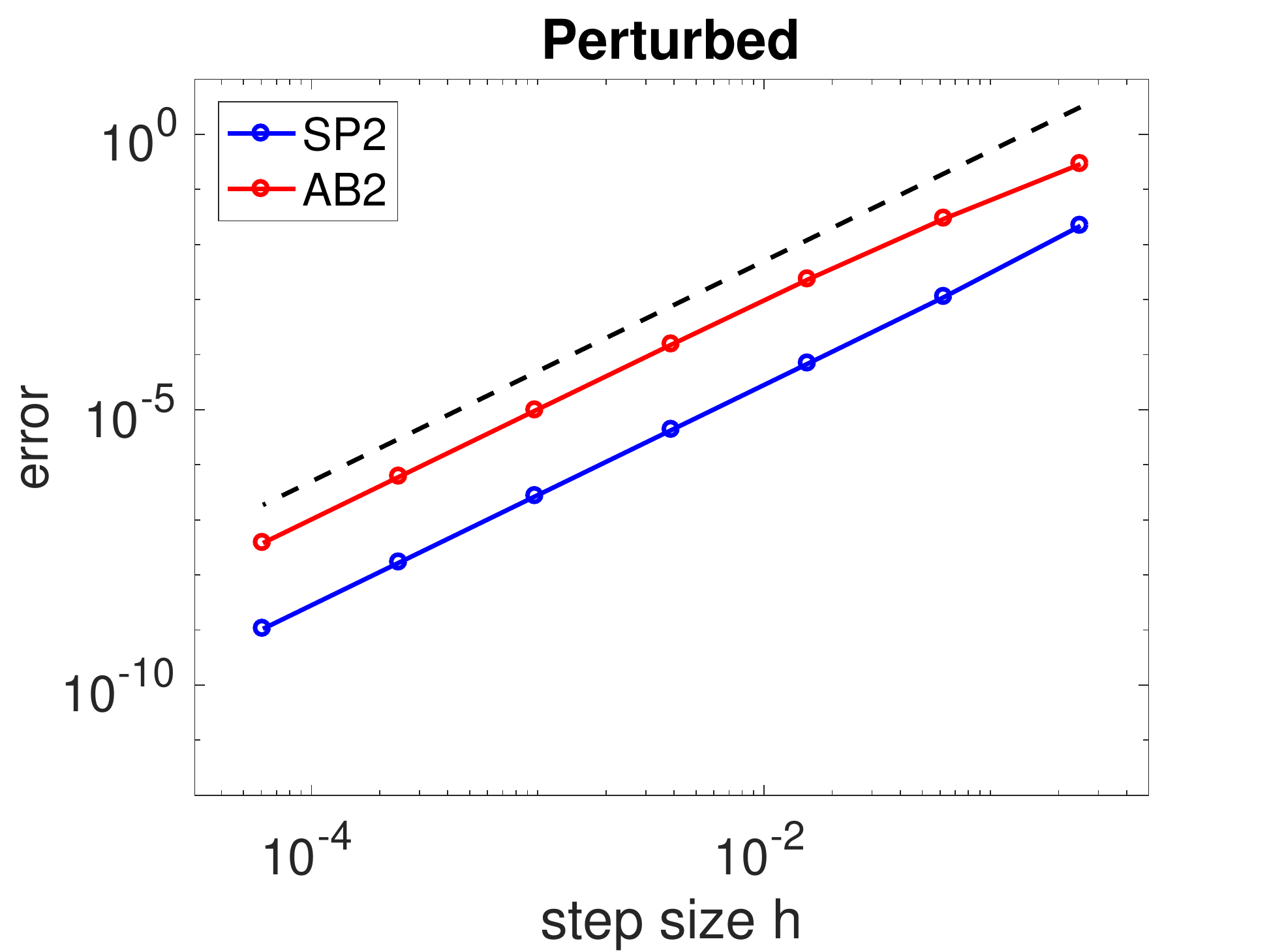}
		\subcaption{Perturbed case.}
		\label{fig:error-h-oblate}
	\end{subfigure}
	~
	\begin{subfigure}{0.45\textwidth}
		\centering
		\includegraphics[width=\textwidth]{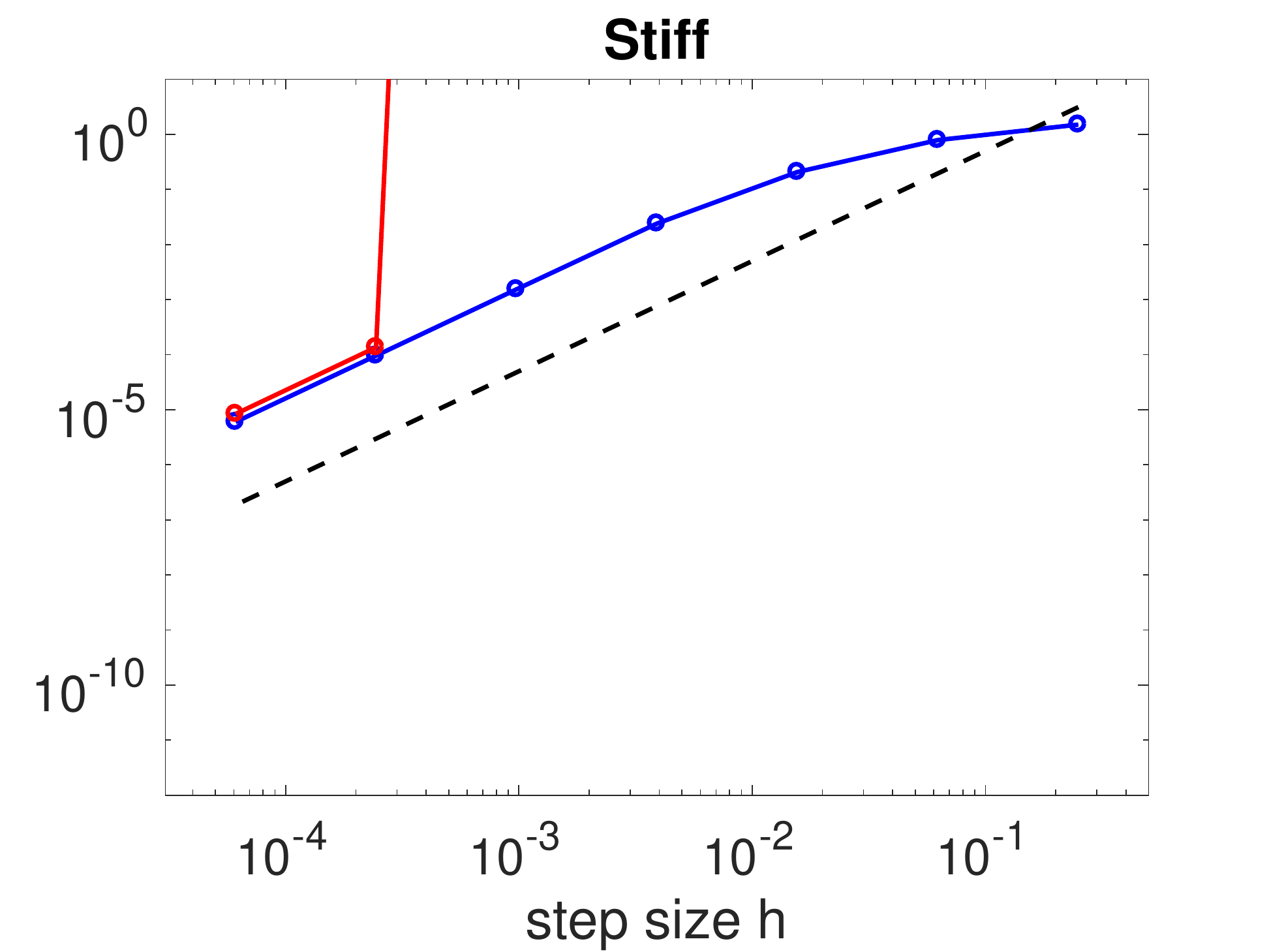}
		\subcaption{Stiff case.}
		\label{fig:error-h-prolate}
	\end{subfigure}
	
	\caption{Second-order convergence of the splitting method (blue line) and the AB2 method (red line).}
	\label{fig:error-h}
\end{figure}

Figure \ref{fig:error-h} shows the second-order convergence of the SP2 solution compared to the AB2 solution for step sizes $h=2^{-n}$ for $n=2,4,6,8,10,12,14$. We observe that both methods achieve the correct order of convergence, however the error of the SP2 solution is significantly lower in the perturbed case compared to the AB2 solution. In the stiff case, the SP2 solution achieves the correct order of convergence for low time-steps and reduced order for larger time-steps. For large time-steps the AB2 solution becomes unstable as denoted by the nearly vertical line. \\

\begin{figure}
	\begin{subfigure}{0.45\textwidth}
		\centering
		\includegraphics[width=\textwidth]{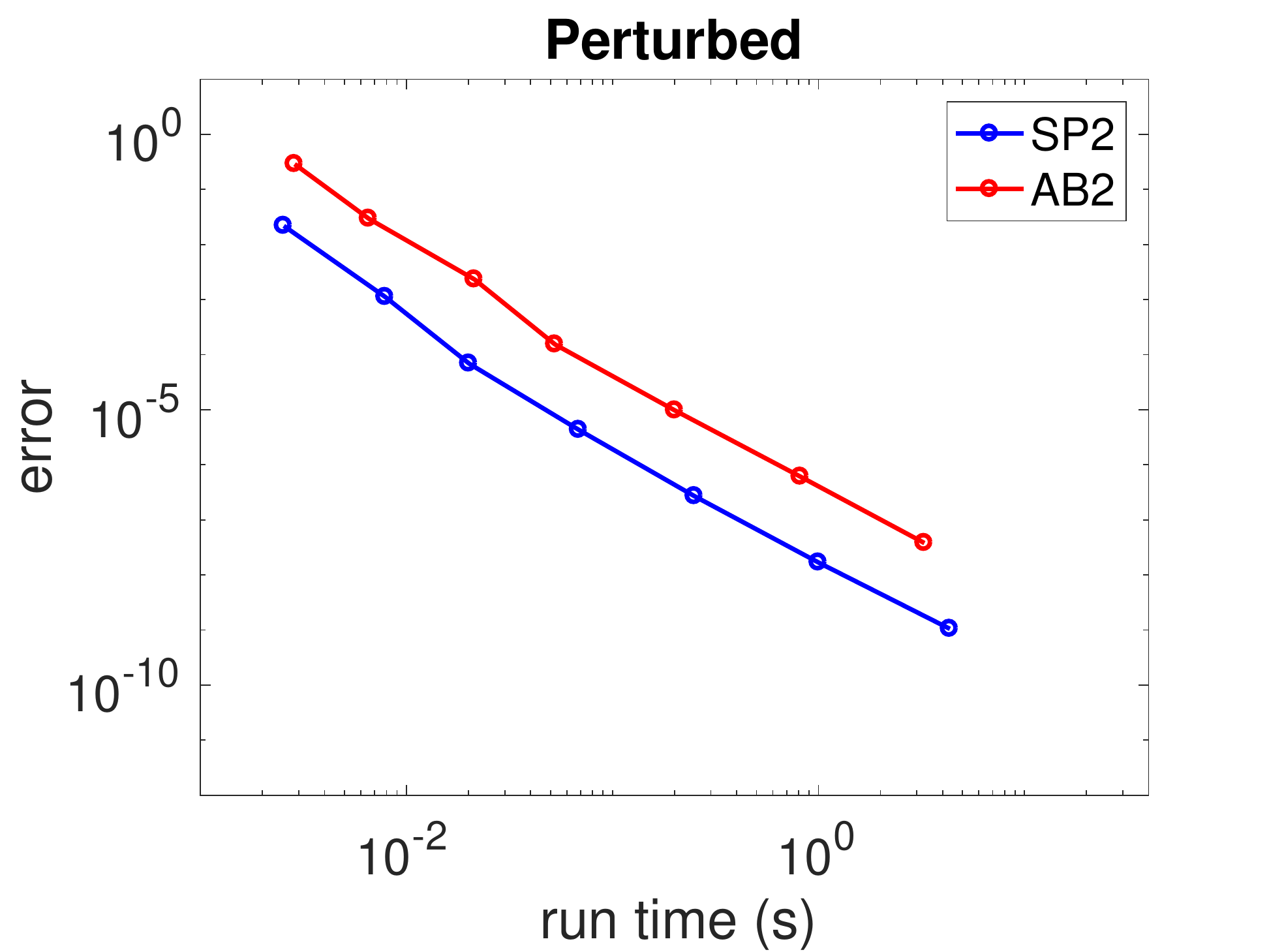}
		\subcaption{Perturbed case.}
		\label{fig:error-cost-oblate}
	\end{subfigure}
	~
	\begin{subfigure}{0.45\textwidth}
		\centering
		\includegraphics[width=\textwidth]{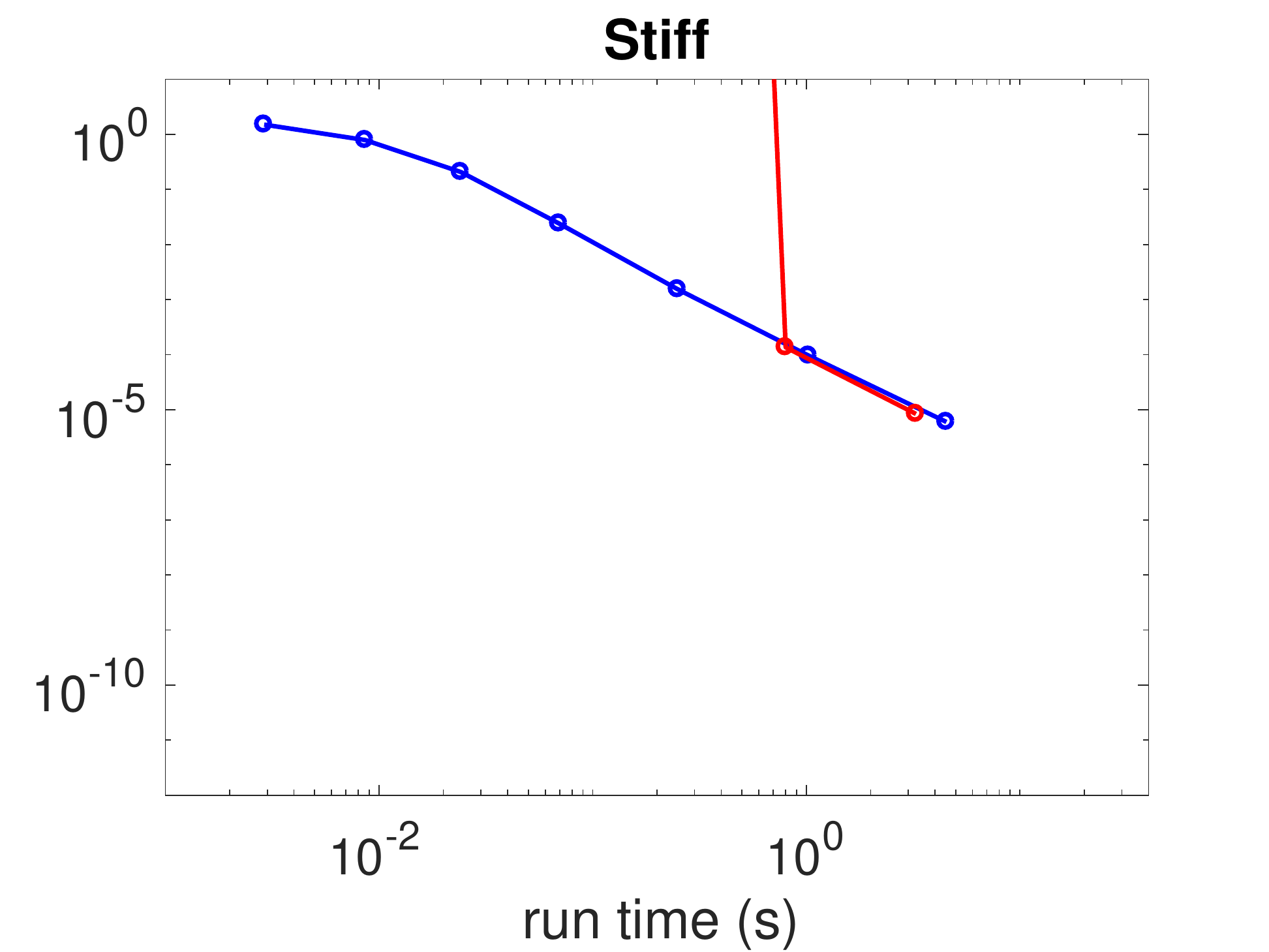}
		\subcaption{Stiff case.}
		\label{fig:error-cost-prolate}
	\end{subfigure}
	
	\caption{Simulation wall-clock time of the splitting method (Blue line) and the AB2 method (red line.}
	\label{fig:error-cost}
\end{figure}

Figure \ref{fig:error-cost} shows the relative computational cost of the two methods measured in simulation wall-clock time for MATLAB serial code implementation. We observe that the SP2 method yields numerical solutions that have over an order of magnitude less error for the same computational cost over the one second interval for the perturbed case.\\

\begin{figure}
	\begin{subfigure}{0.45\textwidth}
		\centering		
		\includegraphics[width=\textwidth]{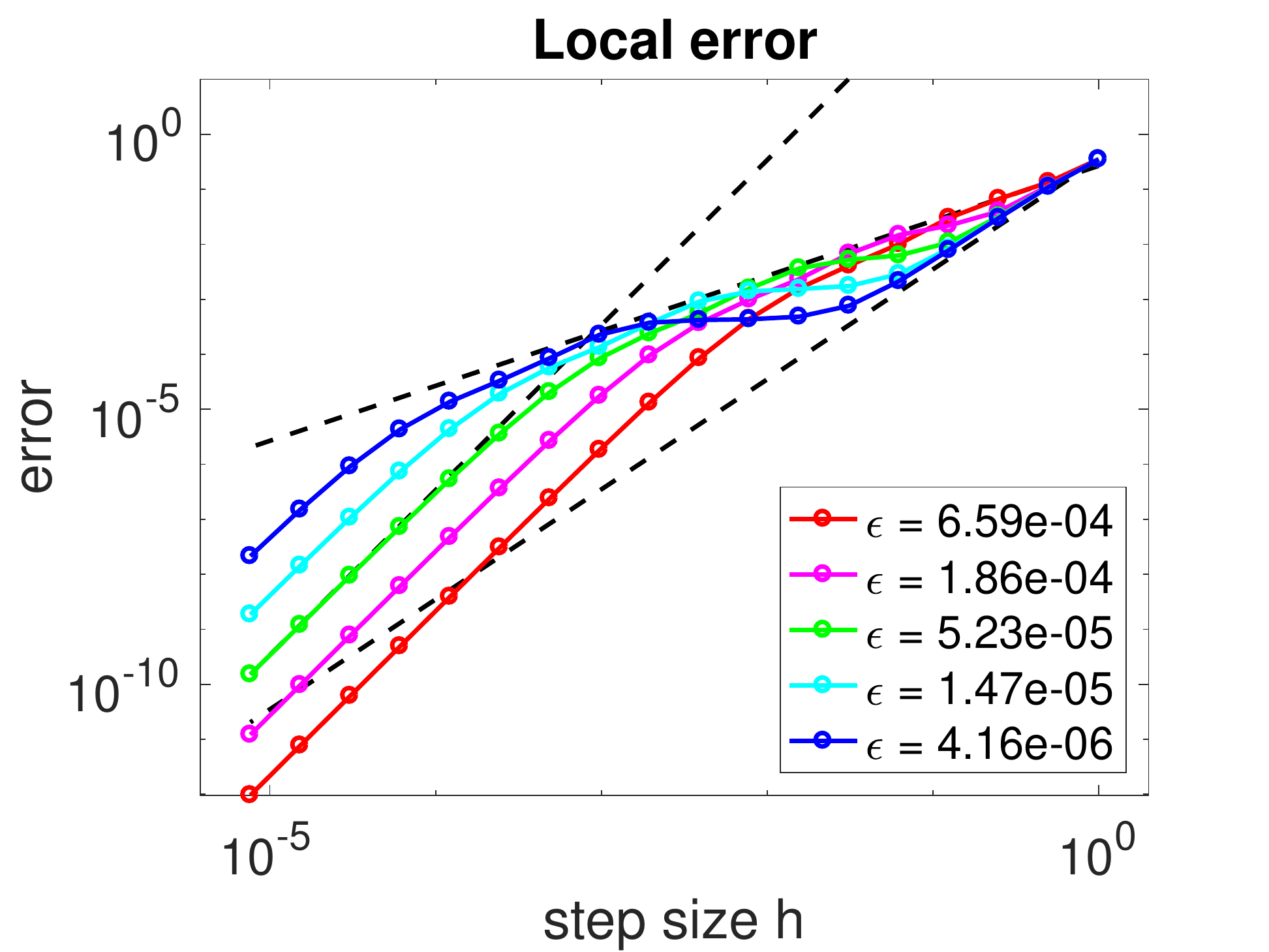}
		\subcaption{}
		\label{fig:stiff_le}
	\end{subfigure}
	~
	\begin{subfigure}{0.45\textwidth}
		\centering
		\includegraphics[width=\textwidth]{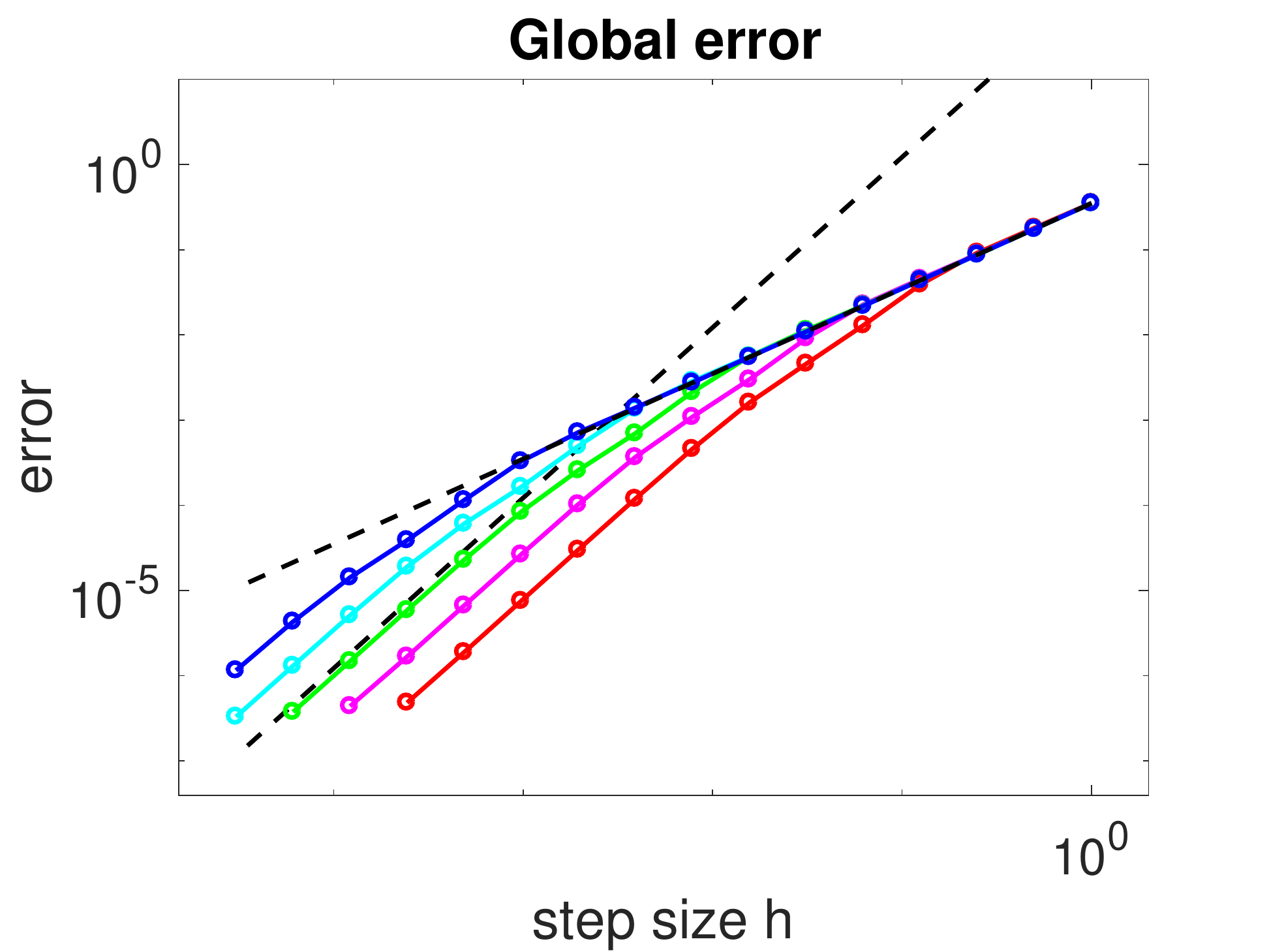}
		\subcaption{}
		\label{fig:stiff_ge}
	\end{subfigure}
	\centering
	
	\caption{Convergence plots for varying stiffness parameters for the local error (a) and global error (b). Order-two and order-one reference lines are plotted on both figures as well as order-three for (a). }\label{fig:stiff_tests}
\end{figure}

\begin{figure} 
	\centering

	\begin{subfigure}{0.45\textwidth}
		\centering		
		\includegraphics[width=\textwidth]{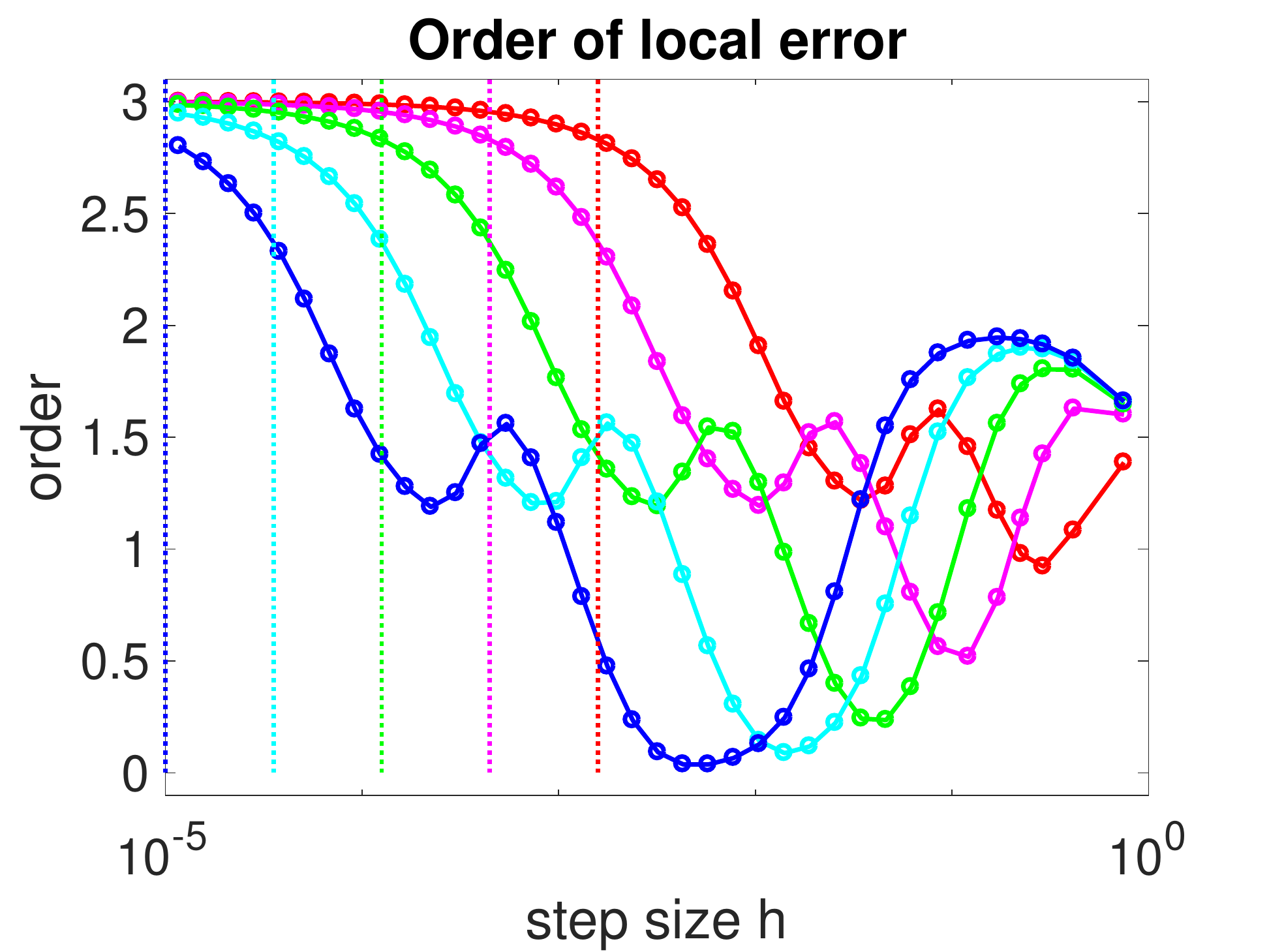}
		\subcaption{}
		\label{fig:stiff_le_order}
	\end{subfigure}
	~
	\begin{subfigure}{0.45\textwidth}
		\centering
		\includegraphics[width=\textwidth]{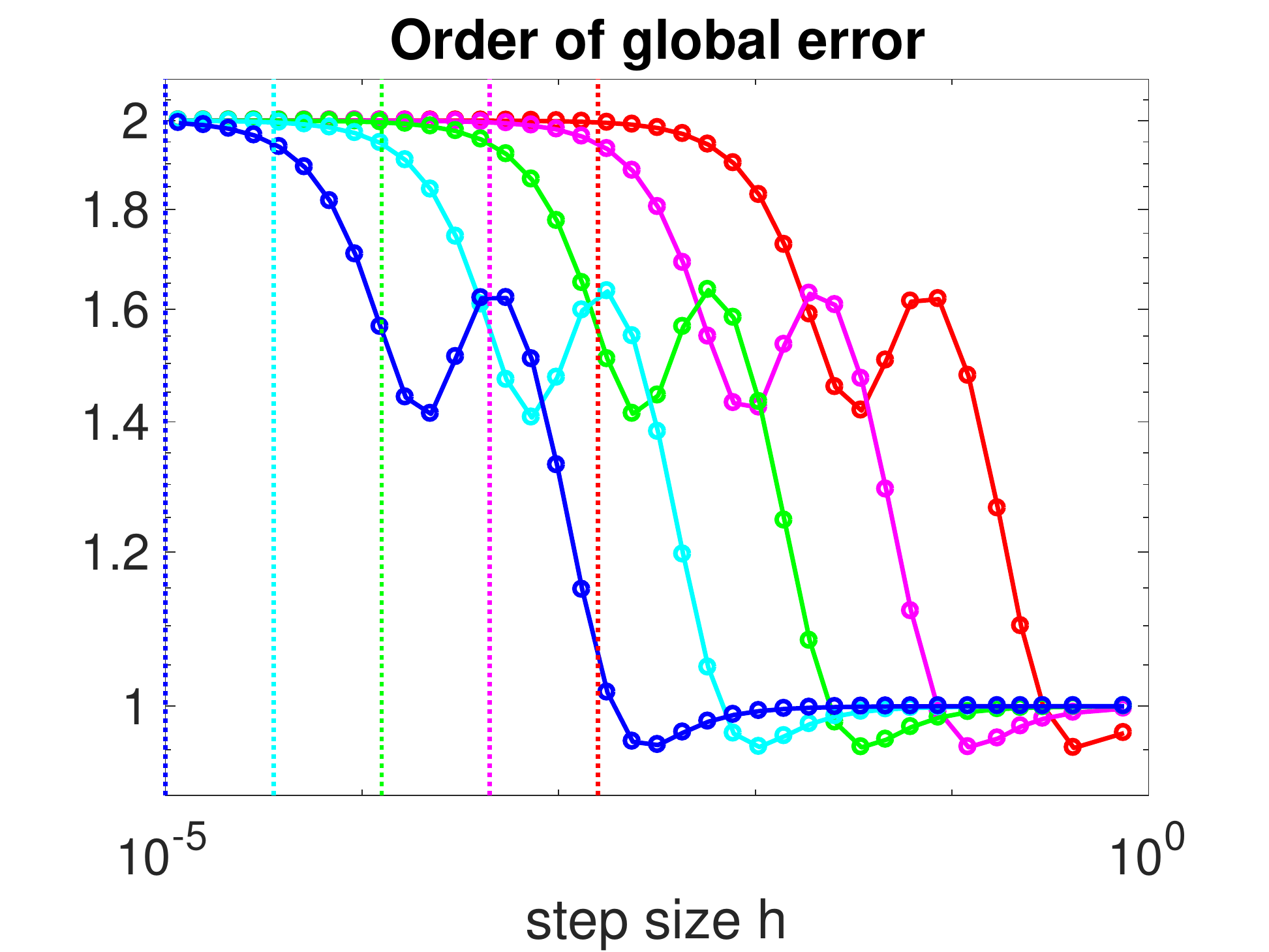}
		\subcaption{}
		\label{fig:stiff_ge_order}
	\end{subfigure}
	\centering
	\caption{The global order for the stiff equation for varying values of $\varepsilon$ (same as figure \ref{fig:stiff_tests}), which are displayed as vertical dotted lines. }\label{fig:stiff_order}
\end{figure}

Figure \ref{fig:stiff_le} shows the local error $||\bd^{[ST]}||$ for varying stiffness parameters $\varepsilon$ that are calculated via $\varepsilon=1/\bar{\gamma}$, where $\bar{\gamma } =||(\gamma^1, \gamma^2, \dots, \gamma^{18})||/18$ for eigenvalues $\gamma^i$ of $A$. Here, we observe the order reduction phenomenon sometimes referred to as the "hump" \cite[p. 113]{SODE2} where we see no increase in error when the step size is increased. This usually occurs in the region $\varepsilon<h<\sqrt{\varepsilon}$ as was observed in \cite{kozlov} for the Van der Pol oscillator when the Strang splitting operator used contains the non-stiff flow operator in the middle. In the non-stiff regime, the local error behaves according to classical theory: it is order-three and proportional to $1/\varepsilon$. In the stiff regime, we observe various order reduction phenomena including convergence to an $\varepsilon$-independent low-order line. In addition to the predictions made by \cite{kozlov}, we observe that the order is also reduced to about 1.5 in the region just below the "hump". This is most clearly observed by the blue line of figure \ref{fig:stiff_le} and is again emphasised in figure \ref{fig:stiff_order}. Figure \ref{fig:stiff_ge} presents the corresponding global errors. As expected, we observe that the solutions are of order two and proportional to $1/\varepsilon$ in the non-stiff regime. As the time-step is increased the order converges to an $\varepsilon$-independent order-one line. Although we perform no rigorous error analysis to explain this, our experiments suggest that there is some $\varepsilon$-independent upper bound of the form $u\leq hc(\by_0)$ for some value $c$ that can depend on the initial conditions $\by_0$. This is highlighted by the dashed order-one reference line. \\ 

Figure \ref{fig:stiff_order} presents the orders of the lines in figure \ref{fig:stiff_tests} and the corresponding values of $\varepsilon$ by vertical dotted lines.  We observe in figure \ref{fig:stiff_le_order} that for $h<\varepsilon$, the method has local order three and as the step size increases, we see some strange $\varepsilon$ dependent order reduction phenomena. The global order of figure \ref{fig:stiff_ge_order} shows a similar phenomenon in the transition region, where the lines go from order two to one in the stiff regime. \\

\begin{figure} [H]
	\centering
	
	\begin{subfigure}{0.45\textwidth}
		\centering
		\includegraphics[width=\textwidth]{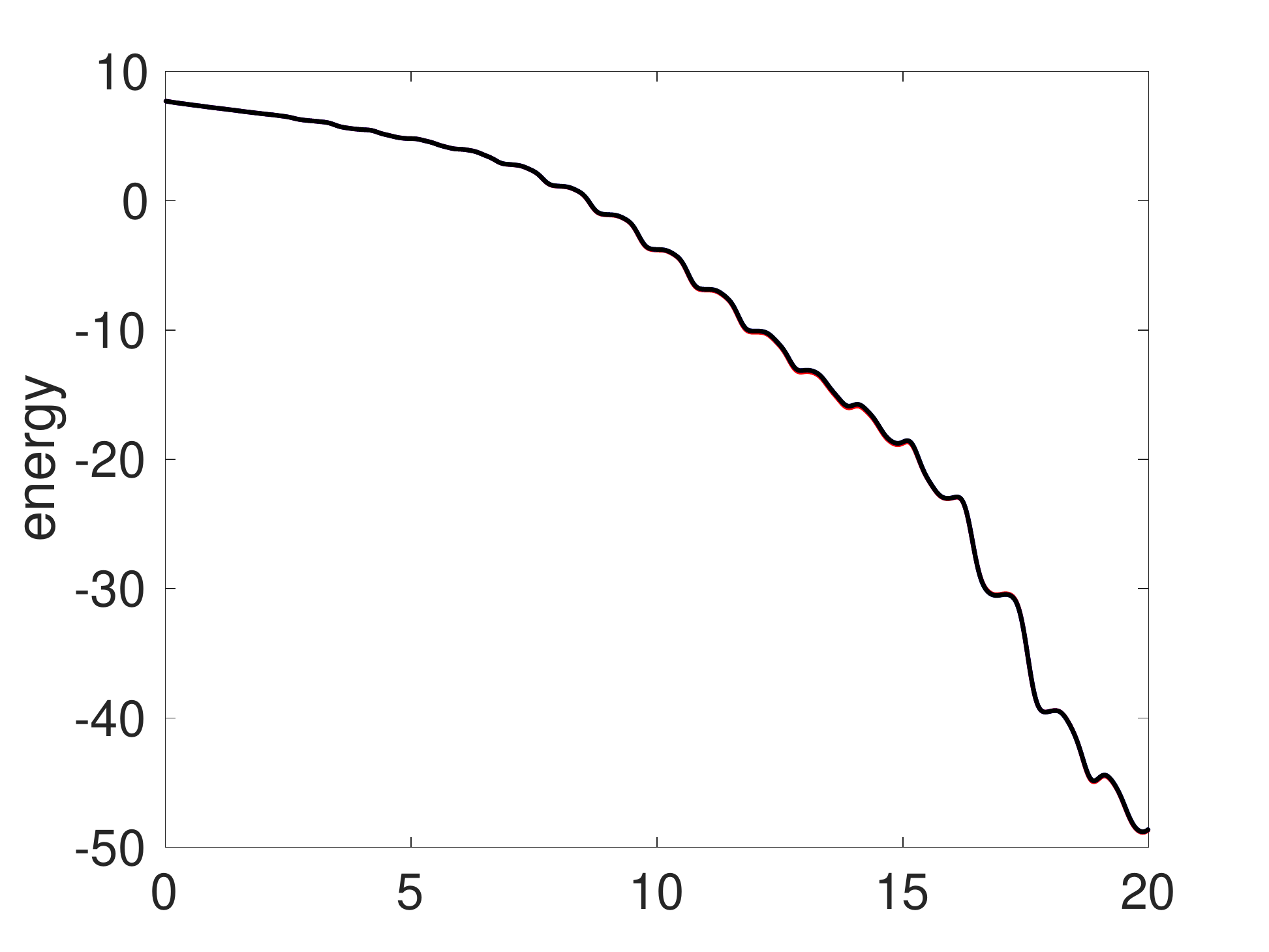}
		\subcaption{}
		\label{fig:energy}
	\end{subfigure}
	~
	\begin{subfigure}{0.45\textwidth}
		\centering
		\includegraphics[width=\textwidth]{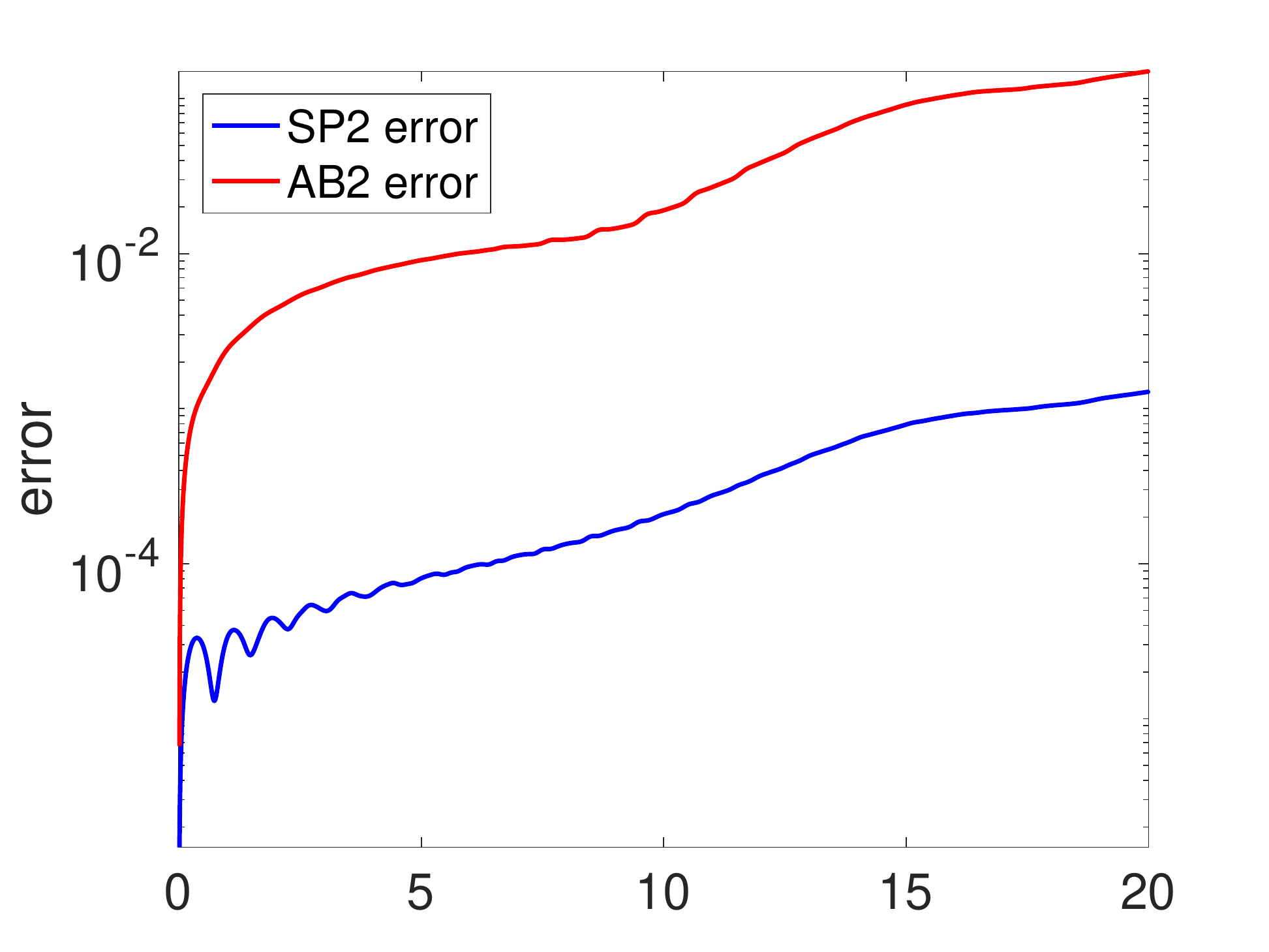}
		\subcaption{}
		\label{fig:error-T}
	\end{subfigure}
	
	\begin{subfigure}{0.45\textwidth}
		\centering
		\includegraphics[width=\textwidth]{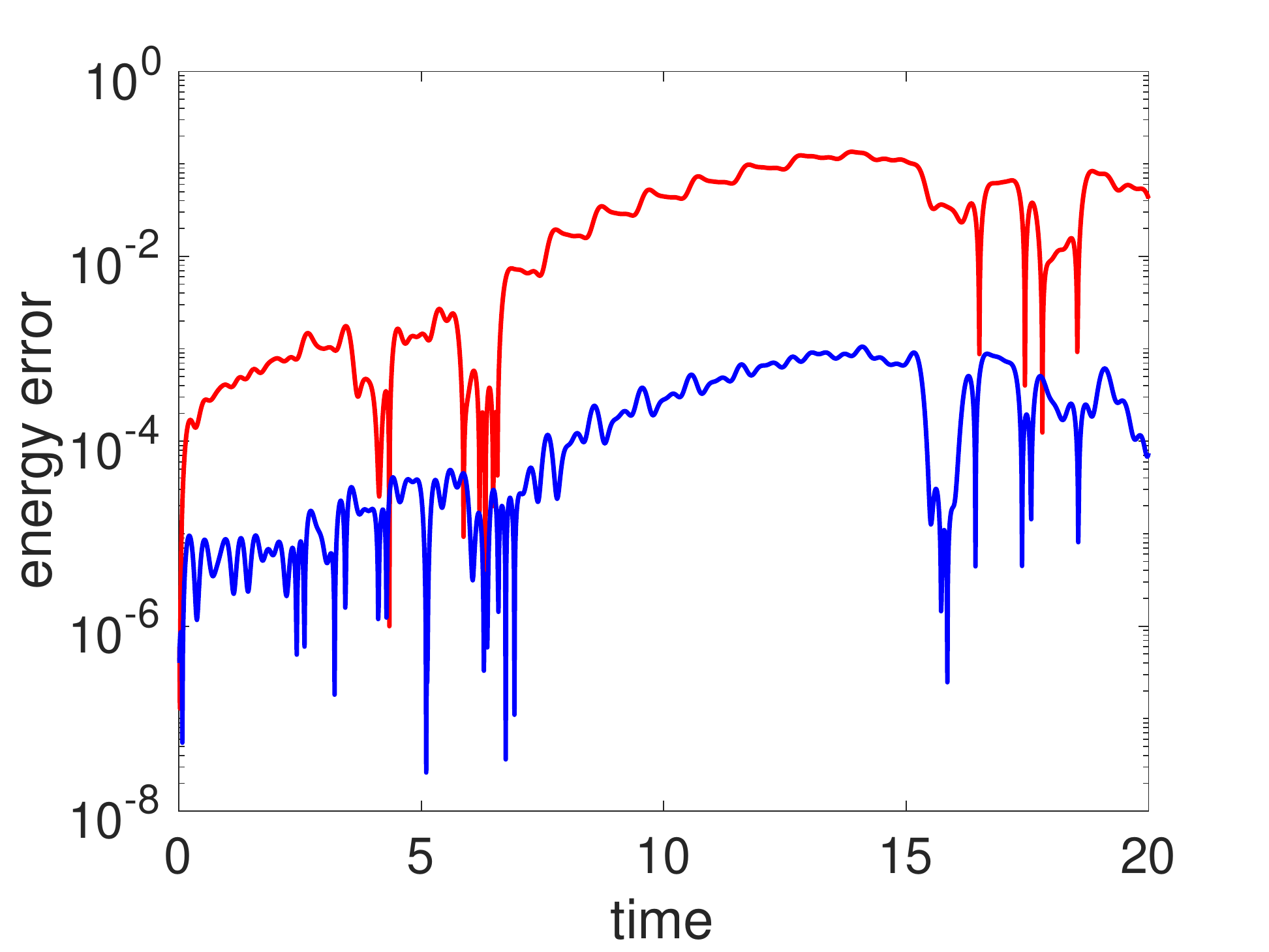}
		\subcaption{}
		\label{fig:energy-error}
	\end{subfigure}
	~
	\begin{subfigure}{0.45\textwidth}
		\centering
		\includegraphics[width=\textwidth]{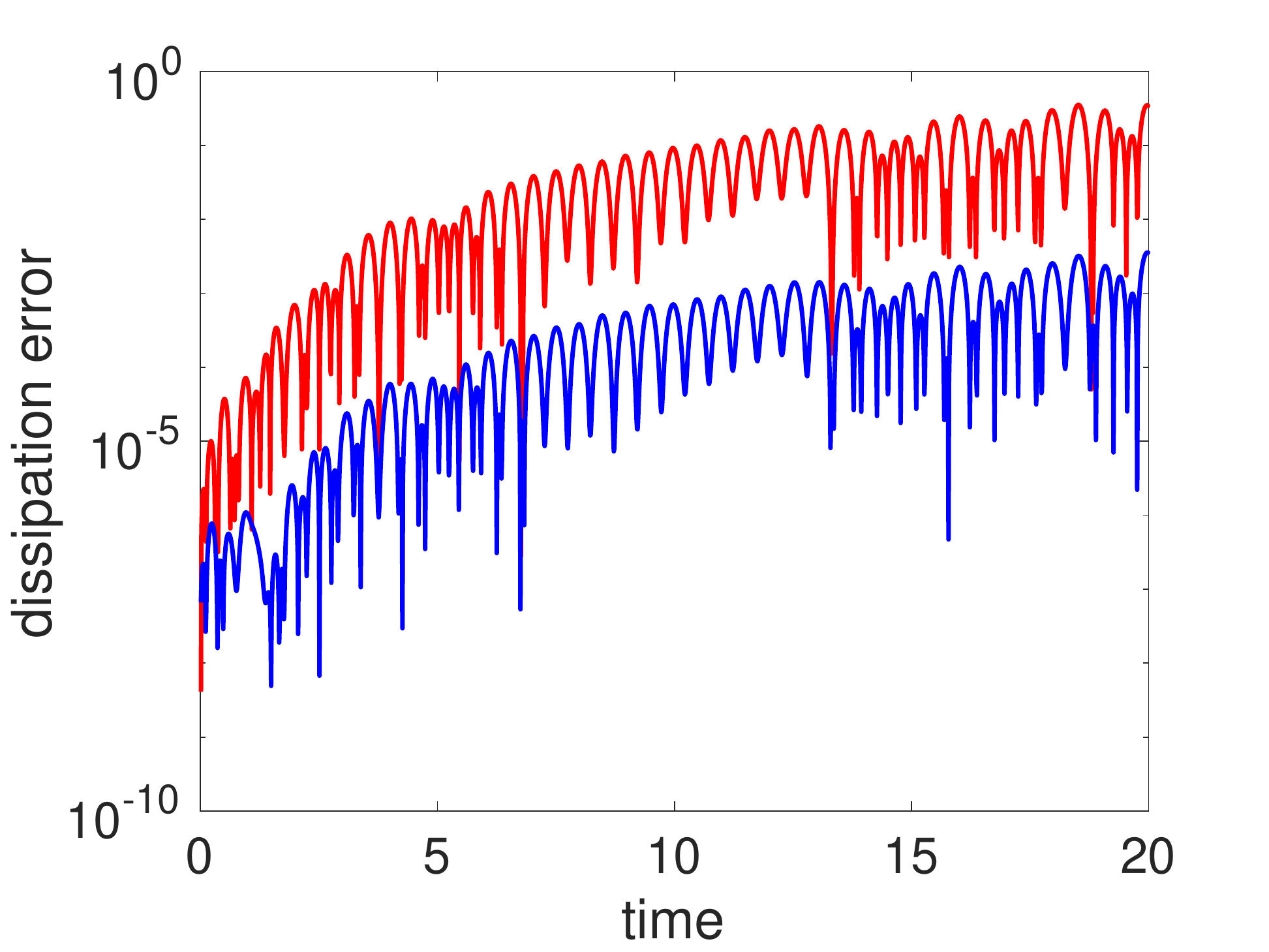}
		\subcaption{}
		\label{fig:energy-dissipation-error}
	\end{subfigure}
	\caption{The particle energy (a), solution error (b), energy error (c) and dissipation error (d) as functions of time for the splitting solution (blue line) and Adams-Bashforth solution (red line) compared to the reference solution (black line) for a perturbed system.}
	\label{fig:energy-dissipation-plots}
\end{figure}

The reference energy $H$ and dissipation $\dot{H}$ are calculated from equations  \eqref{eq:energy} and \eqref{eq:dissipation} using the reference solution and is compared against the numerical energy and dissipation from the SP2 and AB2 solutions in an oscillating shear flow $\mathbf{u}_S$, described in section \ref{ch:fluid}, over a 20 second time interval with time-step $h=0.001$. The system uses the same input parameters as the perturbed case in the previous experiment. Figure \ref{fig:energy} presents the energy of the particle as its dynamics evolves over the 20 second interval. The solution errors are displayed in figure \ref{fig:error-T}, the energy errors are displayed in figure \ref{fig:energy-error} and the dissipation errors are displayed in figure \ref{fig:energy-dissipation-error}, in all cases, the SP2 solution errors are approximately two orders of magnitude lower than those of the AB2.\\

%%%%%%%%%%%%%%%%%%%%%%%%%%%%%%%%%%%%%%%%%%%%%%%%%%%%%%%%%%%%%
%%%%%%%%%%%%%%%%%%%%%%%%%%%%%%%%%%%%%%%%%%%%%%%%%%%%%%%%%%%%%
\section{Conclusion}%%%%%%%%%%%%%%%%%%%%%%%%%%%%%%%%%%%%%%%%%
%%%%%%%%%%%%%%%%%%%%%%%%%%%%%%%%%%%%%%%%%%%%%%%%%%%%%%%%%%%%%
%%%%%%%%%%%%%%%%%%%%%%%%%%%%%%%%%%%%%%%%%%%%%%%%%%%%%%%%%%%%%

We have proposed a splitting method for particle dynamics in viscous flows, obtained by splitting the vector fields of the forced rigid-body dynamics equations into a conservative vector field and a vector field that accounts for the fluid forces. Using backward error analysis, we have shown for perturbed systems, the global error is proportional to $\mathcal{O}(\varepsilon h^2)$ which is an order $\varepsilon$ lower than conventional methods. For the stiff case, the splitting method produces solutions that are stable in the unstable regime of the conventional method and retains stability for all $h\leq1$. Via numerical experiment, we confirm results from the literature \cite{kozlov}, on the local error order reduction phenomena for the splitting method. In the non-stiff regime, the global error is observed to behave according to $\mathcal{O}(h^2/\varepsilon)$ and transitions to $\mathcal{O}(h)$ in the stiff regime.

\section{Acknowledgements}
This work has received funding from the European Unions Horizon 2020 research and innovation
programme under the Marie Sklodowska-Curie grant agreement (No. 691070) as well as the SPIRIT project (No. 231632) under the Research Council of Norway FRIPRO funding scheme. Part of this work was done while visiting the University of Cambridge, UK and La Trobe University, Melbourne, Australia. 
\bibliographystyle{unsrt}
\bibliography{bibliography}{}

\appendix
\section{Local error for Strang splitting}\label{app:strang}
The local error for the Strang operator is given by 
\begin{align}\label{eq:strangle}
	 \bd^{[S]}(\by_0) =& \varphi_h(\by_0)-\Phi^{[S]}_h(\by_0)\nonumber\\
	=& h^3 (\frac{1}{12}[f_1,[f_1,f_2]]-\frac{1}{24}[f_2,[f_2,f_1]])\mathbf{y}_0+\mathcal{O}(h^4),
\end{align} 
and can be computed explicitly by inserting equations \eqref{eq:f1} and \eqref{eq:f2}  
\begin{align}\label{Sle}
	\bd^{[S]}(\by) = \frac{ h^3}{12} \big(&
	-\nabla^2 A (S\nh-\frac{1}{2}A\by,S\nh,\by,\cdot)
	-\nabla A (\mathbf{c}_1,\by,\cdot)
	-\nabla A (2S\nh-\frac{1}{2}A\by,S\nh,\cdot) \nonumber\\
	&+\nabla A (S\nh,\frac{1}{2}A\by,\cdot)
	+\nabla A (\by,S\nabla H-\frac{1}{2}A\by,(\cdot)S\nabla^2H) \nonumber\\ 
	&-\frac{1}{2}\nabla A (S\nh,\by,(\cdot)A)
	+\nabla A (S\nh,\by,(\cdot)\nabla^2HS) \nonumber\\
	&-\nabla S (S\nh-A\by,\nh,(\cdot)A)
	+\nabla S (\mathbf{c}_2,\nh,\cdot)\nonumber\\
	&+\nabla S (S\nh-A\by,\nabla^2HA\by,\cdot)
	+\nabla S (A\by,\nabla^2HS\nh,\cdot)\nonumber\\
	&-\nabla S (A\by,\nh,(\cdot)\nabla S H^2)
	+AS\nabla^2 H(S\nh-A\by)+2S\nabla^2HAS\nh \nonumber\\
	& - S\nabla^2 H S  \nabla^2HA\by 
	- \frac{1}{2}(S\nabla^2HA^2\by+A^2S\nh) \big)
	+\mathcal{O}(h^4), 
\end{align}
 where we have used the fact that the matrix $S$ is linear in $\by$ and vectors $\mathbf{c}_1 = \nabla S (S\nh-A\by,\nh,\cdot)+S\nabla^H(S\nh-A\by) + AS\nh + \nabla A(S\nh,\by,\cdot)$ and $\mathbf{c}_2 = \nabla A(S\nh-\frac{1}{2} A\by,\by,\cdot) + A(S\nh-\frac{1}{2}A\by) + \frac{2}{h^2}\bd^{[LT]}(\by)$ and $A = \varepsilon\tilde{A}$ for the perturbed case. 
\end{document}